\documentclass[12pt]{article}
\usepackage{epsfig,amsmath,amsfonts,amssymb,amstext,afterpage,psfrag,slashbox}
\renewcommand{\arraystretch}{1.2}

\long\def\symbolfootnote[#1]#2{\begingroup%
\def\thefootnote{\fnsymbol{footnote}}\footnote[#1]{#2}\endgroup}

\def\l{\langle}
\def\r{\rangle}

\def\spose#1{\hbox to 0pt{#1\hss}}
\def\lsim{\mathrel{\spose{\lower 3pt\hbox{$\mathchar"218$}}
 \raise 2.0pt\hbox{$\mathchar"13C$}}}
\def\gsim{\mathrel{\spose{\lower 3pt\hbox{$\mathchar"218$}}
 \raise 2.0pt\hbox{$\mathchar"13E$}}}

\catcode`@=11
\def\@citex[#1]#2{%
  \if@filesw\immediate\write\@auxout{\string\citation{#2}}\fi
  \def\@citea{}\@cite{\@for\@citeb:=#2\do
    {\@citea\def\@citea{,\penalty\@m}\@ifundefined
      {b@\@citeb}{{\bf ?}\@warning
{Citation `\@citeb' on page \thepage \space undefined}}%
      \hbox{\csname b@\@citeb\endcsname}}}{#1}}
\def\citer{\@ifnextchar [{\@tempswatrue\@citexr}{\@tempswafalse\@citexr[]}}
  \def\@citexr[#1]#2{%
    \if@filesw\immediate\write\@auxout{\string\citation{#2}}\fi
    \def\@citea{}\@cite{\@for\@citeb:=#2\do
      {\@citea\def\@citea{--\penalty\@m}\@ifundefined
{b@\@citeb}{{\bf ?}\@warning
{Citation `\@citeb' on page \thepage \space undefined}}%
\hbox{\csname b@\@citeb\endcsname}}}{#1}}

\newcommand{\dth}[1]{\frac{d #1}{d \cos \theta}}

\psfrag{costh}{$\cos \theta$}
\psfrag{sqrtsingev}{$\sqrt{s}$ in GeV}
\psfrag{dsiglh}{$\dth{\sigma^{LH}}$ in $R$}
\psfrag{dsigrh}{$\dth{\sigma^{RH}}$ in $R$}

\begin{document}

\begin{titlepage}

\begin{flushright}
{\small
LMU-ASC~01/13\\ 
FLAVOUR(267104)-ERC-34\\
DESY 13-007\\
February 2013\\
%Draft \today
%hep-ph/yymmnnn
}
\end{flushright}

\vspace{0.5cm}
\begin{center}
{\Large\bf \boldmath                                               
Effective Field Theory Analysis of New Physics\\
\vspace*{0.3cm}                                                            
in $e^+e^-\to W^+W^-$ at a Linear Collider
\unboldmath}
\end{center}

\vspace{0.5cm}
\begin{center}
{\sc G. Buchalla$^1$, O. Cat\`a$^1$, R. Rahn$^1$ and M. Schlaffer$^{1,2}$} 
\end{center}

\vspace*{0.4cm}

\begin{center}
$^1$Ludwig-Maximilians-Universit\"at M\"unchen, Fakult\"at f\"ur Physik,\\
Arnold Sommerfeld Center for Theoretical Physics, 
D--80333 M\"unchen, Germany\\
\vspace*{0.2cm}
$^2$DESY, Notkestra\ss e 85, D--22607 Hamburg, Germany
\end{center}

\vspace{1.5cm}
\begin{abstract}
\vspace{0.2cm}\noindent
We analyze new physics contributions to $e^+e^-\to W^+W^-$ at the
TeV energy scale, employing an effective field theory framework.
A complete basis of next-to-leading order operators in the standard model
effective Lagrangian is used, both for the nonlinear and the linear
realization of the electroweak sector. The elimination of redundant
operators via equations-of-motion constraints is discussed in detail.
Polarized cross sections for $e^+e^-\to W^+W^-$ (on-shell) are computed
and the corrections to the standard model results are given in an expansion
for large $s/M^2_W$. The dominant relative corrections grow with $s$ and can
be fully expressed in terms of modified gauge-fermion couplings. 
These corrections are interpreted in the context of the Goldstone boson 
equivalence theorem. Explicit new physics models are considered to 
illustrate the generation and the potential size of the coefficients in the 
effective Lagrangian.
Brief comments are made on the production of $W^+W^-$ pairs at the LHC. 
\end{abstract}

%\vspace*{2.5cm}
%PACS: 11.10.Gh, 11.15.Ex, 12.39.Fe

\vfill
\end{titlepage}

%%%%%%%%%%%%%%%%%%%%%%%%%%%%%%%%%%%%%%%%%%%%%%%%%%%%%%%%%%%%%%%%%
%   Introduction
%%%%%%%%%%%%%%%%%%%%%%%%%%%%%%%%%%%%%%%%%%%%%%%%%%%%%%%%%%%%%%%%%
\section{Introduction}
\label{sec:intro}

During the last decades there has been an intense scrutiny of the 
standard model in the search for traces of new physics effects. However, 
up to the energy scales probed until now and except for some occasional 
tensions, the standard model has proven to be an extremely successful theory. 
Moreover, the latest results from the LHC not only seem to confirm the 
Higgs-like nature of the newly-found 
scalar~\cite{Aad:2012tfa,Chatrchyan:2012ufa} but continuously increase the 
gap between the standard model particles and the scale of new physics 
$\Lambda$. 

In the absence of new heavy particles in direct searches we should expect 
new physics to be first seen as virtual effects. These can generically be 
encoded as anomalous couplings for the different sectors of the theory: 
gauge-fermion interactions, gauge boson interactions 
(oblique, triple-gauge and quartic) and scalar interactions. 
Given the large energy gap between the electroweak and the new physics scale, 
an effective field theory (EFT) treatment becomes the best strategy to 
parametrize the new physics effects in a model-independent way. 
The main virtue of the EFT treatment is that the standard model symmetries 
are automatically implemented in the anomalous couplings. The resulting 
constraints from $SU(2)_L\times U(1)_Y$ symmetry make it transparent that 
(i) the number of independent parameters is typically smaller than the number 
of couplings; (ii) arbitrarily setting some of the couplings to zero 
in experimental analysis is in general inconsistent with the electroweak 
symmetry; (iii) the naive scaling with energy of the form factors is 
ameliorated by ($SU(2)_L\times U(1)_Y$)-induced cancellations. Therefore, 
adopting an EFT becomes not a matter of choice, but the only way to ensure 
consistency at the field theoretical level. The advantages of the EFT
approach have recently been re-emphasized 
in \cite{Degrande:2012wf,Degrande:2013mh}.   
   
Global fits to the electroweak data using an EFT framework have been 
performed by several groups in the past. Unfortunately, the global analysis 
contains too many parameters and cross-correlations are too strong to obtain 
an informative fit~\cite{Han:2004az,Cacciapaglia:2006pk}. As an alternative,
the number of coefficients is commonly 
limited to a reduced set, inspired by the results of different models, and 
fits have been performed on this basis. A prototype of this approach is the 
well-known $S$, $T$, $U$ parameter analysis of \cite{Peskin:1991sw} . 

In this paper we will explicitly show that the above shortcoming of the 
global fit to EFT couplings can be ameliorated by studying individual 
processes at high energies ($v\ll \sqrt{s}\ll \Lambda$). 
As an example, we will 
present a detailed EFT-based study of $W^+W^-$ production at linear 
colliders. $e^+e^-\to W^+W^-$ has been the benchmark process in the study of 
charged triple-gauge corrections first at LEP \cite{Alcaraz:2006mx}
and subsequently for future 
linear collider facilities \citer{Gaemers:1978hg,Andreev:2012cj}. 
%\cite{Gaemers:1978hg,Baur:1987mt,Bilenky:1993ms,Bilenky:1993uy,
%Gounaris:1996rz,Hagiwara:1996kf,AguilarSaavedra:2001rg,Menges:2001gg,
%Weiglein:2004hn,Linssen:2012hp,Andreev:2012cj}
$W^+W^-$ production at the LHC has been considered for instance in
\cite{Weiglein:2004hn,Baur:1988qt,:1999fr,ThurmanKeup:2001ka}. 
However, although there have been 
several studies in the literature emphasizing the need for an EFT approach 
to triple-gauge couplings in $e^+e^-\to W^+W^-$
\cite{Degrande:2012wf,Degrande:2013mh}, \citer{Falk:1991cm,He:1997zm},
%\cite{Falk:1991cm,Bagger:1992vu,Hagiwara:1993ck,He:1997zm}
a complete analysis is still missing. 

The present analysis will be performed in the full nonlinear EFT basis 
recently studied 
in \cite{Buchalla:2012qq}. Our final results will provide expressions 
for the (initial and final state) polarized cross-sections in the large-$s$ 
expansion, which is an excellent approximation for the projected energies at 
future linear colliders. 
The leading corrections will grow with $s$ relative to the standard model
results, reflecting the fact that the nonlinear effective theory
violates unitarity in the UV. The $s/v^2$ enhancement at
energies where the EFT is still valid, improves the visibility of small 
new physics coefficients. Actually, the new physics effects at, say, 
$s=800$~GeV could be typically as large as $20\%$. 

One of the interesting properties of $e^+e^-\to W^+W^-$ is that, up to tiny 
mass corrections, it is independent of couplings to a physical Higgs sector. 
We will show that this is indeed the case by comparing our 
results with the linear EFT basis of 
\cite{Buchmuller:1985jz,Grzadkowski:2010es}. Besides the results for 
the cross-sections, our main findings can be summarized as follows:
\begin{itemize}
\item Despite the sizable number of operators contributing to the process
at next-to-leading order (NLO), 
the final result for new physics effects in the large-$s$ limit
can be encoded in terms of just three parameters. 
These can be expressed as the corrections to the left and 
right-handed gauge-fermion vertices. 
\item Three of the gauge-fermion operators and the three leading 
(C, P and CP-con\-ser\-ving) triple-gauge operators are related by 
field redefinitions. Therefore, in the case of $e^+e^-\to W^+W^-$,
omitting the gauge-fermion operators is not an approximation but 
an exact field-theoretical result: they can be traded for triple-gauge 
operators and vice versa, depending on the chosen operator basis. We stress 
that this is because only three independent gauge-fermion couplings enter
the process $e^+e^-\to W^+W^-$. In general, there are many more
gauge-fermion operators and it is not possible to eliminate all of them.
\end{itemize} 
The last point above implies that statements about gauge-fermion or 
triple-gauge operators {\it{per se}} are basis-dependent and therefore 
ill-defined. For instance, in the basis where gauge-fermion operators are 
kept, the electroweak fit~\cite{Han:2004az} does not support the common 
claim that 
they are tightly constrained. Furthermore, our analysis contradicts the 
statement that $W^+W^-$ production directly tests triple-gauge corrections. 
Rather, what one finds is that at large-$s$ one can put bounds on 
gauge-fermion couplings or {\it{equivalently}} on triple-gauge couplings, 
since they are not independent.   

The existence of field-theoretical relations binding gauge-fermion, 
triple-gauge and oblique operators raises the question of which basis should 
be preferred for experimental analyses of electroweak physics. 
In the particular case of $e^+e^-\to W^+W^-$ the possibility of eliminating 
the gauge-fermion operators altogether might suggest itself. However, in view 
of the general electroweak fit, it seems more natural to eliminate triple-gauge 
operators and keep the full set of gauge-fermion operators. As we will show, 
the emerging picture in this basis turns out to be rather simple: only a 
single triple-gauge operator appears (${\cal O}_{XU3}$ in (\ref{oxu}) below), 
which is both parity and isospin-breaking, and therefore expected to be 
numerically small.\footnote{CP-violating triple-gauge operators are also 
present but do not interfere with the standard model in the cross sections.} 
Additionally, 
in the large-$s$ limit oblique corrections and the surviving triple-gauge 
operator can be shown to be generically subleading, such that the leading 
large-$s$ contribution naturally singles out gauge-fermion operators.   

These rather simple and counterintuitive results follow from carefully 
eliminating redundant operators and therefore stress the importance of 
working with a complete and minimal basis in EFT-based analyses. 
Comments on how this picture would generalize to hadron colliders 
will be made but details will be left to future work.  

This paper is organized as follows: in section~2 we will briefly review 
the EFT of the standard model at NLO and fix our notation and conventions. 
In section~3 
we will apply the EFT formalism to $e^+e^-\to W^+W^-$, discussing in detail 
direct contributions and parameter redefinitions. In section~4 we collect the 
results for the differential cross sections for the different initial and 
final state polarizations. The issue of redundant operators and 
choice of basis is addressed in section~5. A complementary view of the 
large-$s$ limit from the perspective of the equivalence theorem 
is given in section~6. 
In section~7 we discuss the case of a linearly-realized EFT.
In order to get an estimate of the expected effects at linear 
colliders, in section~8 the size of EFT couplings is estimated 
from different benchmark UV completions.
In section~9 we briefly comment on $W^+W^-$ production at the LHC. 
Conclusions are given in section~10, while technical details are relegated 
to an Appendix.  

%%%%%%%%%%%%%%%%%%%%%%%%%%%%%%%%%%%%%%%%%%%%%%%%%%%%%%%%%%%%%%%%%

%\afterpage{\clearpage}

%%%%%%%%%%%%%%%%%%%%%%%%%%%%%%%%%%%%%%%%%%%%%%%%%%%%%%%%%%%%%%%%%        
%   U-basis                                                    
%%%%%%%%%%%%%%%%%%%%%%%%%%%%%%%%%%%%%%%%%%%%%%%%%%%%%%%%%%%%%%%%%       
  
\section{Electroweak chiral Lagrangian at NLO}
\label{sec:nlou}

The starting point of our analysis is the well-known leading order
chiral Lagrangian of the electroweak standard model. 
To define our notation we quote here the terms of the leptonic
sector relevant for $e^+e^-\to W^+W^-$. They read
\begin{equation}\label{lsmlo}
{\cal L}_{\rm LO} = -\frac{1}{2}\langle W_{\mu\nu}W^{\mu\nu}\rangle 
-\frac{1}{4} B_{\mu\nu}B^{\mu\nu} + \frac{v^2}{4}\ \l D_\mu U^\dagger D^\mu U\r
+\bar l_L i\!\not\!\! Dl_L + \bar e_R i\!\not\!\! De_R 
\end{equation}
Here and in the following the trace of a matrix $M$ is written as
$\langle M\rangle$.
The doublet of left-handed leptons is denoted by $l_L=(\nu_L,e_L)^T$, 
the right-handed electron by $e_R$, and we focus
our attention on the first-generation fermions. 
The covariant derivatives of the fermions are 
\begin{equation}\label{dcovf}
D_\mu l_L =
\partial_\mu l_L +i g W_\mu l_L  -\frac{i}{2} g'  B_\mu l_L , 
\qquad
D_\mu e_R =\partial_\mu e_R  - i g' B_\mu e_R
\end{equation}
The electron mass is negligible and the associated Yukawa terms
have been omitted from (\ref{lsmlo}). Couplings to a physical Higgs
field do not play a role in $e^+e^-\to W^+W^-$ and are likewise omitted
from the Lagrangian. The Goldstone 
bosons of electroweak symmetry breaking are represented by the matrix field 
\begin{equation}\label{uudef}
U=\exp(2i\Phi/v),\qquad
\Phi=\varphi^a T^a=\frac{1}{\sqrt{2}}\left(
\begin{array}{cc}
\frac{\varphi^0}{\sqrt{2}} & \varphi^+\\
\varphi^- & -\frac{\varphi^0}{\sqrt{2}} 
\end{array}\right)
\end{equation}
with $T^a=T_a$ the generators of $SU(2)$. The $U$-field transforms as 
\begin{equation}\label{uglgr}
U\rightarrow g_L U g^\dagger_R,\qquad g_{L,R}\in SU(2)_{L,R}
\end{equation}
where $g_L$ and the $U(1)_Y$ subgroup of $g_R$
are gauged, so that the covariant derivative of $U$ is given by
\begin{equation}\label{dcovu}
D_\mu U=\partial_\mu U+i g W_\mu U -i g' B_\mu U T_3
\end{equation}

The effective Lagrangian (\ref{lsmlo}) describes 
physics at the electroweak scale $v=246\,{\rm GeV}$, assumed to be 
small in comparison with a new physics scale $\Lambda$. 
This Lagrangian is non-renormalizable in general, except
when a Higgs field $h$ is introduced with specific couplings,
in which case the theory reduces to the conventional standard model
(see e.g. \cite{Contino:2010rs} for a review). In the general case,
additional terms will arise beyond the lowest order from
the dynamics of electroweak symmetry breaking at the ${\rm TeV}$ scale.
These subleading terms were first considered in
\citer{Longhitano:1980iz,Appelquist:1994qz}.  
%\cite{Longhitano:1980iz,Longhitano:1980tm,
%Appelquist:1984rr,Hauser:1985ey,Appelquist:1993ka,Appelquist:1994qz}.
A complete list of all NLO operators in this framework
based on a systematic 
power counting has recently been given in \cite{Buchalla:2012qq}.
Using the notation of this paper, the NLO operators relevant
for $e^+e^-\to W^+W^-$ can be written as
\begin{equation}\label{lnloco}                                              
{\cal L}_{\rm NLO} =\beta_1 {\cal O}_{\beta_1} + 
\sum_{i=1}^{6} C_{Xi} {\cal O}_{XUi} + 
\sum_{i=7}^{10} C_{Vi} {\cal O}_{\psi Vi} + 
C^*_{V9} {\cal O}^\dagger_{\psi V9} + \frac{C_{4f}}{\Lambda^2} {\cal O}_{4f} +
\sum_{i=1}^{2} \frac{C_{Wi}}{\Lambda^2} {\cal O}_{Wi}
\end{equation}
with operators ${\cal O}_k$ specified in (\ref{ob1}) -- (\ref{owi}).
The complete basis of NLO operators \cite{Buchalla:2012qq} also
contains the terms $\bar e_L e_R W^+_\mu W^{-\mu}$ and 
$\bar e_L\sigma^{\mu\nu} e_R W^+_\mu W^-_\nu$, which could in principle
contribute to $e^+e^-\to W^+W^-$. Due to the chirality flip in the electron
current the coefficients of these operators can be expected to be
proportional to the Yukawa coupling of the electron and thus
very much suppressed. In addition,
the chirality-changing currents do not interfere with the vectorial currents
of the leading-order amplitude. Those operators therefore give no
first-order correction to the $e^+e^-\to W^+W^-$ cross sections  
and we have omitted them from (\ref{lnloco}).
We have included the 4-fermion operator ${\cal O}_{4f}$, which contributes
only indirectly through the renormalization of the Fermi constant $G_F$.
Other 4-fermion operators from \cite{Buchalla:2012qq} do not give rise to
first-order corrections to $e^+e^-\to W^+W^-$ cross sections and have been
neglected.

The operators ${\cal O}_{Wi}$ (see (\ref{owi}) below) are strictly
speaking terms that appear only at next-to-next-to-leading order (NNLO) in the 
effective Lagrangian. Their coefficients are generally loop-induced
\cite{Arzt:1994gp} and count as $C_{Wi}\sim 1/(16\pi^2)\sim v^2/\Lambda^2$, 
which multiplies the explicit prefactor $1/\Lambda^2$ in the last term
of (\ref{lnloco}). We have included them here in order to
facilitate the transition to the basis of operators within the framework 
of a linearly transforming Higgs field, to be considered in 
section~\ref{sec:nlophi}.
In this case they belong to the full list of operators of dimension 6,
and we include them for completeness in our analysis.
In the present context, and working consistently to NLO,
the coefficients $C_{Wi}$ may be put to zero.

All operators in the Lagrangian (\ref{lnloco}) are hermitian and have real 
coefficients, except ${\cal O}_{\psi V9}$. They have already been known from 
the work of \cite{Longhitano:1980tm,Appelquist:1984rr,Appelquist:1993ka}. 
However, the basis of operators used there contains redundant terms,
which can be eliminated using the equations of motion 
\cite{Buchalla:2012qq,Nyffeler:1999ap,Grojean:2006nn}.

The operators in (\ref{lnloco}) have the following explicit form, where the 
second expression in each case refers to unitary gauge with $U=1$:
\begin{equation}\label{ob1}
{\cal O}_{\beta_1} = v^2 \langle U^\dagger D_\mu U T_3\rangle^2
= -M^2_Z Z_\mu Z^\mu  
\end{equation}

\begin{eqnarray}\label{oxu}  
{\cal O}_{XU1} &=& g'g\ B_{\mu\nu}\ \l U^\dagger W^{\mu\nu} U T_3\r 
                =\frac{g'g}{2}\ B^{\mu\nu}W_{\mu\nu}^{3} \nonumber\\
{\cal O}_{XU2} &=& g^2\ \l U^\dagger W_{\mu\nu} U T_3\r\
   \l U^\dagger W^{\mu\nu} U T_3\r 
  =\frac{g^2}{4}W_{\mu\nu}^{3}W^{3\mu\nu}\nonumber\\
{\cal O}_{XU3} &=& g\ \varepsilon^{\mu\nu\lambda\rho}\
   \l U^\dagger W_{\mu\nu} D_\lambda U\r\ 
   \l U^\dagger D_\rho U T_3\r \nonumber\\
   &=& \frac{g}{4}\varepsilon^{\mu\nu\lambda\rho}
\left[gW_{\mu\nu}^{a}W_{\lambda}^{a}-g'W_{\mu\nu}^{3}B_{\lambda}\right]
\left[g'B_{\rho}-gW_{\rho}^{3}\right]  \nonumber\\
{\cal O}_{XU4} &=& g' g\ \varepsilon^{\mu\nu\lambda\rho}\
  B_{\mu\nu}\ \l U^\dagger W_{\lambda\rho} U T_3\r          
    = \frac{g'g}{2}\varepsilon^{\mu\nu\lambda\rho}
     B_{\mu\nu}W_{\lambda\rho}^{3} \nonumber\\
{\cal O}_{XU5} &=& g^2\ \varepsilon^{\mu\nu\lambda\rho}\ 
\l U^\dagger W_{\mu\nu} U T_3\r\ \l U^\dagger W_{\lambda\rho} U T_3\r  
   = \frac{g^2}{4}\varepsilon^{\mu\nu\lambda\rho}
      W_{\mu\nu}^{3}W_{\lambda\rho}^{3}\nonumber\\
{\cal O}_{XU6} &=& g\ \l U^\dagger W_{\mu\nu} D^\mu U\r\ 
\l U^\dagger D^\nu U T_3\r \nonumber\\
  &=& \frac{g}{4} \left[gW_{\mu\nu}^{a}W^{a\mu}-
    g' W_{\mu\nu}^{3}B^{\mu}\right]\left[g' B^{\nu} - g W^{3\nu} \right]
\end{eqnarray}

\begin{eqnarray}\label{opsiv}
{\cal O}_{\psi V7} &=& \bar l_L\gamma^\mu l_L\ \l U^\dagger iD_\mu UT_3\r 
=-\displaystyle\frac{\sqrt{g^2+g'^2}}{2}\bar l_L\gamma^\mu l_L Z_{\mu} \nonumber\\
{\cal O}_{\psi V8} &=& \bar l_L\gamma^\mu UT_3U^\dagger l_L\ 
\l U^\dagger iD_\mu UT_3\r 
=-\displaystyle\frac{\sqrt{g^2+g'^2}}{2}\bar l_L\gamma^\mu T_3 l_L Z_{\mu}
\nonumber\\
{\cal O}_{\psi V9} &=& \bar l_L\gamma^\mu U P_{12} U^\dagger l_L\ 
\l U^\dagger iD_\mu U P_{21}\r 
=-\displaystyle\frac{g}{\sqrt{2}} \bar\nu_L \gamma^{\mu} e_L W_{\mu}^+ \nonumber\\
{\cal O}_{\psi V10} &=& \bar e_R\gamma^\mu e_R\ \l U^\dagger iD_\mu UT_3\r
=-\displaystyle\frac{\sqrt{g^2+g'^2}}{2}\bar e_R\gamma^\mu e_R  Z_{\mu}
\end{eqnarray}

\begin{eqnarray}\label{op4f}
{\cal O}_{4f} &=& \frac{1}{2} ({\cal O}_{LL5}- 4 {\cal O}_{LL15})
=\bar e_L\gamma^\mu\mu_L\, \bar\nu_{\mu L}\gamma_\mu\nu_{e L} + h.c.
\end{eqnarray}
where the appropriate flavour structure is understood for
${\cal O}_{LL5}$, ${\cal O}_{LL15}$ from \cite{Buchalla:2012qq}.

In (\ref{opsiv}) we have used the definitions $P_{12}\equiv T_1 + i T_2$,
$P_{21}\equiv T_1 - i T_2$. 
It is convenient to work with the following linear combinations
of operators ${\cal O}_{\psi V7,8}$
\begin{equation}\label{opsivpm}
{\cal O}_{\psi V\pm}\equiv \frac{1}{2}{\cal O}_{\psi V7}\pm {\cal O}_{\psi V8}
\end{equation}
whose coefficients become
\begin{equation}\label{cvpm}
C_{V\pm}\equiv C_{V7}\pm \frac{1}{2} C_{V8}
\end{equation}
Only one of these coefficients, $C_{V-}$, appears in the amplitudes for 
$e^+e^-\to W^+W^-$ within our approximations. This is most clearly seen in 
unitary gauge, where ${\cal O}_{\psi V-}$ couples the $Z$ to 
electrons and ${\cal O}_{\psi V+}$ to neutrinos.

Finally, the NNLO terms ${\cal O}_{Wi}$ are
\begin{eqnarray}\label{owi}
{\cal O}_{W1} &=& g^3 \varepsilon^{abc} 
              W^{a\nu}_\mu  W^{b\lambda}_\nu  W^{c\mu}_\lambda  \nonumber\\
{\cal O}_{W2} &=& g^3 \varepsilon^{abc} 
              \tilde W^{a\nu}_\mu  W^{b\lambda}_\nu  W^{c\mu}_\lambda 
\end{eqnarray}
with 
\begin{equation}\label{wtilde}
\tilde W^a_{\mu\nu}=\frac{1}{2}\varepsilon_{\mu\nu\rho\sigma} W^{a,\rho\sigma}\, ,
\qquad \varepsilon^{0123}=-1
\end{equation}

%%%%%%%%%%%%%%%%%%%%%%%%%%%%%%%%%%%%%%%%%%%%%%%%%%%%%%%%%%%%%%%%%           
%   anomalous couplings                                                     
%%%%%%%%%%%%%%%%%%%%%%%%%%%%%%%%%%%%%%%%%%%%%%%%%%%%%%%%%%%%%%%%%           
\section{Anomalous couplings}
\label{sec:ancoup}

The NLO terms in the effective Lagrangian modify the
lowest order vertices of the standard model. Their effect
can be cast in the form of anomalous couplings.

For the triple-gauge vertex (TGV), coupling a virtual, neutral vector boson
$V$ to a $W^+W^-$ pair in the final state, the Feynman rule can be written as 
\begin{equation}\label{tgvgam1}
V^\rho(k)\to W^{-\mu}(p)W^{+\nu}(q)\, :\qquad 
-i \left\{\begin{array}{c} g c_Z\\ g s_Z\end{array}\right\}
\Gamma^{\mu\nu\rho}_V(p,q;k)\, ,\quad 
V=\left\{\begin{array}{c} Z\\ A \end{array}\right.
\end{equation}
where \cite{Appelquist:1993ka,Hagiwara:1986vm} 
\begin{eqnarray}\label{tgvgam2}
\Gamma^{\mu\nu\rho}_V(p,q;k) &=& g^V_1 (p-q)^\rho g^{\mu\nu} +
(g^V_1+\kappa_V)(k^\mu g^{\nu\rho}-k^\nu g^{\mu\rho}) \nonumber\\
&& +i g^V_4 (k^\mu g^{\nu\rho}+k^\nu g^{\mu\rho}) 
-i g^V_5 \varepsilon^{\mu\nu\lambda\rho} (p-q)_\lambda 
+\tilde\kappa_V \varepsilon^{\mu\nu\lambda\rho} k_\lambda \nonumber\\
&& -\frac{\lambda_V}{\Lambda^2} (p-q)^\rho k^\mu k^\nu
-\frac{\tilde\lambda_V}{\Lambda^2} (p-q)^\rho 
\varepsilon^{\mu\nu\sigma\tau} p_\sigma q_\tau
\end{eqnarray}
Here $s_Z$, $c_Z$ are, respectively, sine and cosine of the weak
mixing angle in the $Z$-standard definition ($\alpha=\alpha(M_Z)$)
\begin{equation}\label{szcz}
s^2_Z c^2_Z\equiv\frac{\pi\alpha}{\sqrt{2} G_F M^2_Z}
\end{equation}
and $g$ is the $SU(2)_L$ gauge coupling, where  
$g s_Z = e=\sqrt{4\pi\alpha}$. The anomalous-coupling parameters
in (\ref{tgvgam2}) encode deviations from the standard model, in which 
$g^V_1=\kappa_V=1$ and $g^V_{4,5}=\tilde\kappa_V=\lambda_V=\tilde\lambda_V=0$. 

Similarly, the gauge-fermion interactions can be parametrized through
the Feynman rules
\begin{equation}\label{gfvert}
\begin{array}{rcccc}
\bar\nu e_{L,R}W\, : & \hspace*{1cm} & -\frac{ig}{\sqrt{2}}\kappa_c\gamma^\mu P_L 
                    & \hspace*{1cm} & 0 \\
\bar e e_{L,R}Z\, : & & \frac{ig}{2c_Z}(\kappa_1-2 s^2_Z\kappa_2)\gamma^\mu P_L 
                   &  & \frac{ig}{2c_Z}(-2 s^2_Z\kappa_2)\gamma^\mu P_R 
\end{array}
\end{equation}
for left- and right-handed electrons, respectively, with the
corresponding projectors $P_{L,R}=(1\mp\gamma_5)/2$.
The couplings to the photon ($\bar ee_{L,R}A$) are not modified 
by anomalous couplings because of electromagnetic gauge invariance. 
The $\kappa_i$ in (\ref{gfvert}) parametrize deviations from the standard
model, in which $\kappa_c=\kappa_1=\kappa_2=1$. 

Working in the framework of an effective theory, the anomalous couplings
should be expressed in terms of the operator coefficients in the
effective Lagrangian (\ref{lnloco}). 
The operators ${\cal O}_{\beta_1}$, ${\cal O}_{XU1}$ and ${\cal O}_{XU2}$
contain terms bilinear in the gauge fields $Z$ and $A$, which
can be absorbed into the canonical kinetic terms through the 
renormalizations \cite{Holdom:1990xq} (see also \cite{Appelquist:1993ka})
\begin{equation}\label{zaren}
Z_0=(1+\delta_Z)Z,\qquad A_0=(1+\delta_A)A + \delta_{AZ} Z,\qquad
M_{Z0}=(1-\delta_{M_Z})M_Z
\end{equation}
Here the subscript $0$ denotes fields and parameters in the absence
of any NLO terms in the Lagrangian. We also have
\begin{equation}\label{gse0}
g_0 s_0=e_0,\qquad e_0=(1-\delta_A)e,\qquad G_{F0}=(1-2\delta_G)G_F
\end{equation}
and, from (\ref{szcz}),
\begin{equation}\label{s0c0}
s_0 c_0 = s_Z c_Z (1-\delta_A+\delta_{M_Z}+\delta_G)
\end{equation}

Corrections to the Fermi constant come from ${\cal O}_{V9}$ and
${\cal O}_{4f}$. They lead to
\begin{equation}\label{deltag}
\delta_G=\frac{1}{2} {\rm Re}~(C^e_{V9}+C^\mu_{V9})-
    \frac{v^2}{4\Lambda^2} C_{4f} =C_{V9} - \frac{v^2}{4\Lambda^2} C_{4f}
\end{equation}
The first expression allows for general, flavour non-universal and 
complex coefficients of ${\cal O}_{V9}$. In the opposite case, $\delta_G$ 
simplifies to the second expression in (\ref{deltag}).  

Within the basis of operators in (\ref{ob1}) -- (\ref{opsiv}) the
anomalous couplings can finally be expressed as
\begin{equation}\label{gvkap1}
g^Z_1 = 1+\left[\frac{\beta_1-\delta_G +C_{X1} e^2/c^2_Z}{c^2_Z-s^2_Z}\right]
+3\frac{e^2}{s^2_Z}\frac{k^2}{\Lambda^2}C_{W1}, 
\qquad  g^A_1 = 1 +3\frac{e^2}{s^2_Z}\frac{k^2}{\Lambda^2}C_{W1}
\end{equation}
\begin{equation}\label{gvkap2}
\kappa_Z = 
1+\left[\frac{\beta_1-\delta_G +C_{X1} e^2/c^2_Z}{c^2_Z-s^2_Z}\right] +
\frac{e^2}{c^2_Z} C_{X1}-\frac{e^2}{s^2_Z} C_{X2}
+ 3\frac{e^2}{s^2_Z}\frac{2M^2_W -k^2}{\Lambda^2}C_{W1}
\end{equation}
\begin{equation}\label{gvkap3}
\kappa_A = 1-\frac{e^2}{s^2_Z}(C_{X1}+C_{X2})
+ 3\frac{e^2}{s^2_Z}\frac{2M^2_W -k^2}{\Lambda^2}C_{W1}
\end{equation}
\begin{equation}\label{gvkap4}
g^Z_4 = \frac{e^2}{4 s^2_Z c^2_Z} C_{X6}, \qquad  g^A_4 =0
\end{equation}
\begin{equation}\label{gvkap5}
g^Z_5 = -\frac{e^2}{2 s^2_Z c^2_Z}C_{X3},\qquad  g^A_5 =0 
\end{equation}
\begin{equation}\label{gvkap6}
\tilde\kappa_Z = 2\left(\frac{e^2}{c^2_Z}C_{X4}-\frac{e^2}{s^2_Z}C_{X5}\right)
-6\frac{e^2}{s^2_Z}\frac{M^2_W}{\Lambda^2}C_{W2}
\end{equation}
\begin{equation}\label{gvkap7}
\tilde\kappa_A = -2\frac{e^2}{s^2_Z}(C_{X4}+C_{X5})
-6\frac{e^2}{s^2_Z}\frac{M^2_W}{\Lambda^2}C_{W2}
\end{equation}
\begin{equation}\label{gvkap8}
\lambda_{Z,A} = 6\frac{e^2}{s^2_Z} C_{W1}, \qquad  
\tilde\lambda_{Z,A} =  6\frac{e^2}{s^2_Z} C_{W2}
\end{equation}

\begin{eqnarray}
\kappa_c &=& 1 +\left[\frac{C_{X1}e^2 + c^2_Z(\beta_1-\delta_G)}{c^2_Z-s^2_Z}-
\frac{C_{X2} e^2}{2 s^2_Z}\right] + C_{V9}\\
\kappa_1 &=& 1+\left[\beta_1 -\delta_G\right] -C_{V-}+C_{V10}\\
\kappa_2 &=& 1+\left[\frac{\delta_G-\beta_1-C_{X1} e^2/s^2_Z}{c^2_Z-s^2_Z}\right]
              +\frac{1}{2s^2_Z} C_{V10}\label{kapc12}
\end{eqnarray}

The terms in (\ref{gvkap1}) through (\ref{kapc12}) that arise from
renormalizing $A$, $Z$, $e$, $s_Z$, $c_Z$ are indicated by square brackets.
The remaining corrections represent the direct effect of the NLO operators
on the interaction vertices.
Note that the coefficients $\beta_1$ and $\delta_G$ always appear
in the combination $\beta_1 - \delta_G$.

%%%%%%%%%%%%%%%%%%%%%%%%%%%%%%%%%%%%%%%%%%%%%%%%%%%%%%%%%%%%%%%%%           
%   cross sections                                                          
%%%%%%%%%%%%%%%%%%%%%%%%%%%%%%%%%%%%%%%%%%%%%%%%%%%%%%%%%%%%%%%%%           
\section{Cross sections}
\label{sec:crosec}

In the following we present cross-section formulas for
$e^+e^-\to W^+W^-$, focussing on the new-physics corrections from
the NLO Lagrangian (\ref{lnloco}). 
The amplitude is determined by the $s$-channel ($Z$, $\gamma$) and
$t$-channel ($\nu$) exchange diagrams. Of particular interest for a future 
linear collider will be the limit of large centre-of-mass energy $\sqrt{s}$, 
defined as $v^2 \ll s\ll \Lambda^2$ \cite{Passarino:2012cb}. In this window 
$\sqrt{s}$ is considered to be much larger than the electroweak 
scale $v$, and also $M_{W,Z}$, but still smaller than the new-physics scale 
$\Lambda$ that determines the range of validity of the effective theory.
With the inequality $M^2_{W,Z} \ll s$, the corrections to the
cross sections can be expanded in inverse powers of $s$. Relative to the 
standard model, the potentially leading corrections grow as ${\cal O}(s)$,
subleading terms are of ${\cal O}(1)$, whereas all further terms, suppressed
as ${\cal O}(v^2/s)$ or higher, can be expected to be irrelevant in practice.

We provide results for cross sections with different polarizations of the 
initial and final state particles \cite{Ahn:1988fx}. The case of left-handed 
and right-handed $e^-$ will be denoted by $LH$ and $RH$, respectively. 
For the $W^+W^-$ bosons we consider either longitudinal ($L$) or transverse 
polarization ($T$) for each, which leads to the cases $LL$, $TT$ and $LT$.
For $LT$ the cross sections are the same whether $W^+$ or $W^-$ is 
longitudinally polarized ($LT=TL$).
The polarized cross sections are quoted relative to their standard model
expressions, where only the ${\cal O}(s)$ enhanced terms are given here.
The corrections of ${\cal O}(1)$, $f^{LH}_{LL},\ldots$, 
can be found in the appendix.

\begin{figure}[htbp]
\centering
\includegraphics[width=12cm]{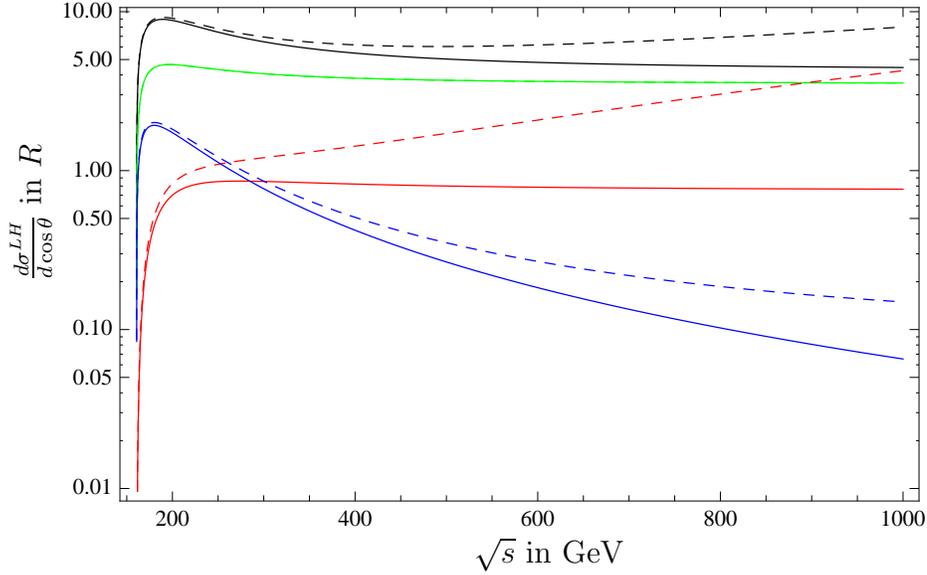}
\caption{Energy dependence of scattering cross sections for left-handed 
electrons at $\cos\theta=0$ in 
units of $R=4\pi\alpha^2/3s$. The solid curves are, from top to bottom 
(at $\sqrt{s}=600\,{\rm GeV}$), the leading-order standard model results for 
unpolarized $W^+W^-$, and for $W$ polarizations $TT$, $LL$ and $LT$.
The dashed curves are the corresponding results including leading  
new physics corrections. Note that these corrections are absent in the 
$TT$ case.}\label{fig:lhs}
\end{figure}

\begin{figure}[htbp]
\centering
\includegraphics[width=12cm]{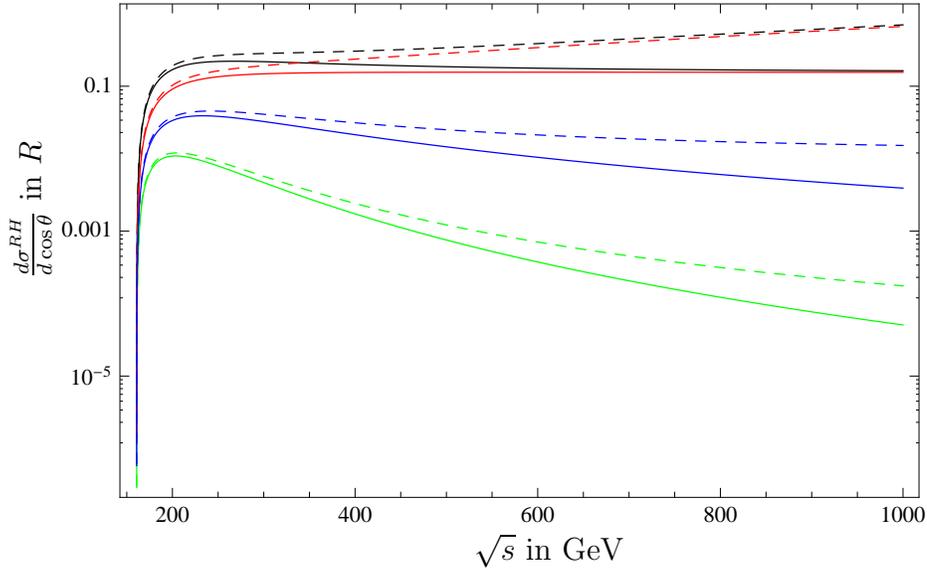}
\caption{Energy dependence of scattering cross sections for right-handed 
electrons at $\cos\theta=0$ in units of $R$. 
The solid curves are, from top to bottom 
(at $\sqrt{s}=600\,{\rm GeV}$), the leading-order standard model results for 
unpolarized $W^+W^-$, and for $W$ polarizations $LL$, $LT$ and $TT$.
The dashed curves are the corresponding results including leading  
new physics corrections.}\label{fig:rhs}
\end{figure}

\begin{figure}[htbp]
\centering
\includegraphics[width=12cm]{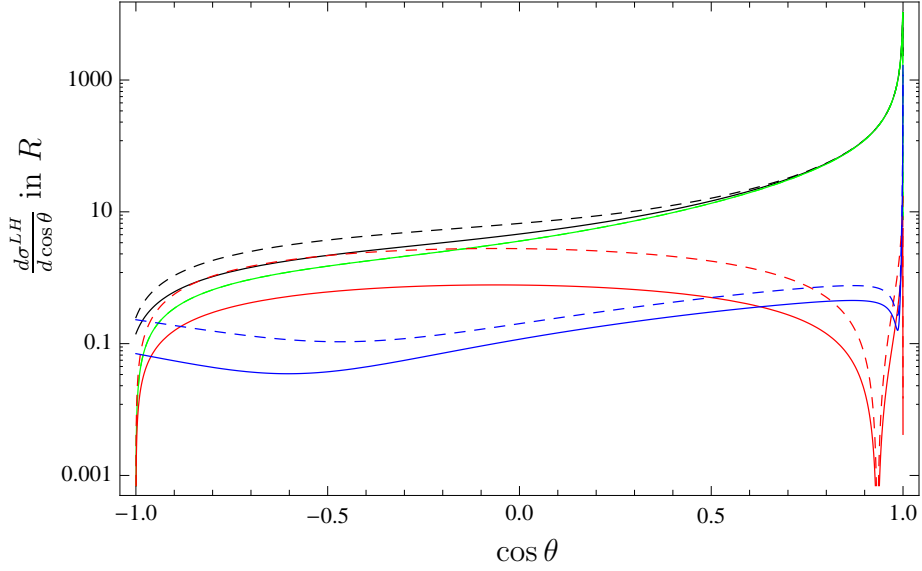}
\caption{Angular dependence of scattering cross sections for left-handed 
electrons at $s=(750\,{\rm GeV})^2$ in units of $R$. 
The solid curves are, from top to bottom 
(at $\cos\theta=0$), the leading-order standard model results for 
unpolarized $W^+W^-$, and for $W$ polarizations $TT$, $LL$ and $LT$.
The dashed curves are the corresponding results including leading  
new physics corrections. Note that these corrections are absent in the 
$TT$ case.}\label{fig:lha}
\end{figure}

\begin{figure}[htbp]
\centering
\includegraphics[width=12cm]{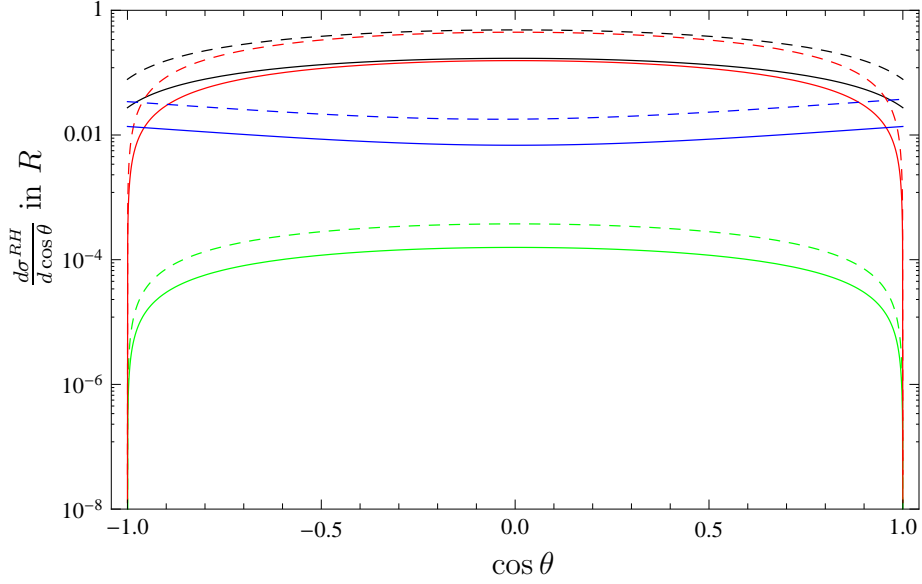}
\caption{Angular dependence of scattering cross sections for right-handed 
electrons at $s=(750\,{\rm GeV})^2$ in units of $R$. 
The solid curves are, from top to bottom 
(at $\cos\theta=0$), the leading-order standard model results for 
unpolarized $W^+W^-$, and for $W$ polarizations $LL$, $LT$ and $TT$.
The dashed curves are the corresponding results including leading  
new physics corrections.}\label{fig:rha}
\end{figure}

The $LL$ cross sections read:

\begin{align}                                                                 
\begin{aligned}\label{slhll}
\dth{\sigma_{LL}^{LH}} &= \dth{\sigma_{LL, SM}^{LH}} \bigg[ 1 + 
s \frac{4 \operatorname{Re} C_{ V9}^e}{M_Z^2} + s \frac{2 C_{V-}}{M_Z^2} + 
f_{LL}^{LH} + \mathcal{O} \left( s^{-1} \right) \bigg]                          
\end{aligned}                                                                 
\end{align}

\begin{align}                                                                 
\begin{aligned}\label{srhll}               
\dth{\sigma_{LL}^{RH}} &= \dth{\sigma_{LL, SM}^{RH}} \bigg[ 1 + 
s \frac{C_{ V10}}{M_Z^2 s_Z^2} + f_{LL}^{RH} + 
\mathcal{O} \left( s^{-1} \right) \bigg]                                       
\end{aligned}                                                               
\end{align}

The notation ${\rm Re}~C_{ V9}^e$ reflects the fact that, in general,
the coefficient $C_{V9}$ may be complex and flavour dependent.
If these possibilities are neglected ${\rm Re}~C_{ V9}^e$ can be
identified with $C_{V9}$ (taken to be real), as it is frequently
done throughout this paper.

The $TT$ cross sections read:

\begin{align}                                                                 
\begin{aligned}\label{slhtt}     
\dth{\sigma_{TT}^{LH}} &= \dth{\sigma_{TT, SM}^{LH}} \bigg[ 1 + f_{TT}^{LH} +
 \mathcal{O} \left( s^{-1} \right) \bigg]                                     
\end{aligned}                                                                 
\end{align}

\begin{align}                                                                 
  \begin{aligned}\label{srhtt}            
 \dth{\sigma_{TT}^{RH}} &= \dth{\sigma_{TT, SM}^{RH}} \bigg[ 1 + 
 s \frac{C_{ V10}}{M_Z^2 s_Z^2} - s \frac{2 e^2 C_{X1}}{M_Z^2 s_Z^2 c_Z^2} +
 s\frac{C_{W1}}{\Lambda^2} \frac{6 e^2}{s_Z^2}+ f_{TT}^{RH} \bigg]              
  \end{aligned}                                                               
\end{align}
In (\ref{srhtt}) the $s$-dependence of the square bracket is exact,
$f^{RH}_{TT}=0$, and terms of ${\cal O}(s^{-1})$ or higher are absent
in this case.

The $LT$ cross sections are given by:

\begin{align}                                                                 
  \begin{aligned}\label{slhlt}                       
\dth{\sigma_{LT}^{LH}} &= \dth{\sigma_{LT, SM}^{LH}} \bigg[ 1 +
s\frac{4 \operatorname{Re}C_{ V9}^e \xi}{M_Z^2 \chi} + 
s\frac{2C_{V-} \xi}{M_Z^2 \chi} -s\frac{e^2 \xi C_{X1}}{M_Z^2 c_Z^2 \chi} - 
s\frac{e^2 \xi C_{X2}}{M_Z^2 s_Z^2 \chi}\\                                     
&-s \frac{e^2 \left( c_Z^2 - s_Z^2 \right) \left[ \left( 1+ \cos \theta 
\right) c_Z^2 + \cos \theta \right] C_{X3} }{M_Z^2 s_Z^2 c_Z^2 \chi} - 
s \frac{C_{W1}}{\Lambda^2}\frac{6 e^2 c_Z^2 \xi}{s_Z^2\chi} \\                
& + f_{LT}^{LH} + \mathcal{O} \left( s^{-1} \right) \bigg]              
  \end{aligned}                                                               
\end{align}

\begin{align}                                                                 
\begin{aligned}\label{srhlt}                
\dth{\sigma_{LT}^{RH}} &= \dth{\sigma_{LT, SM}^{RH}} \bigg[ 1 - 
s \frac{e^2 C_{X3} \cos \theta}{M_Z^2 s_Z^2 c_Z^2 (1 + \cos^2 \theta)} -
s\frac{e^2 C_{X1}}{M_Z^2 c_Z^2 s_Z^2} +s\frac{C_{ V10}}{M_Z^2 s_Z^2} \\    
& \quad + f_{LT}^{RH} + \mathcal{O} \left( s^{-1} \right) \bigg]              
\end{aligned}                                                                 
\end{align}

with

\begin{align}                                                                 
  \begin{aligned}                                                             
    \xi &= 1+\left( 2 c_Z^2 \left(1+ \cos \theta \right) + 
\cos \theta \right) \cos \theta \\                                      
  \chi &= 1+ \left( 2 c_Z^2 \left(1+ \cos \theta \right) + 
\cos \theta \right)^2 \\                                                        
  \end{aligned}                                                             
\end{align}

Finally, we give the corresponding results also for the case
of unpolarized $W$ bosons (denoted by $\Sigma$):

\begin{align}                                                                 
\begin{aligned}\label{slhsum}                                              
\dth{\sigma_{\Sigma}^{LH}} &= \dth{\sigma_{\Sigma, SM}^{LH}} \bigg[ 1 + 
s \frac{16 \operatorname{Re} C_{ V9}^e \sin^4 \frac{\theta}{2}}{M_Z^2 \eta } 
+ s \frac{8 C_{V-} \sin^4 \frac{\theta}{2}}{M_Z^2 \eta } + f_{\Sigma}^{LH} + 
\mathcal{O} \left( s^{-1} \right) \bigg]                                    
\end{aligned}                                                                 
\end{align}

\begin{align}                                                                 
\begin{aligned}\label{srhsum}                                       
\dth{\sigma_{\Sigma}^{RH}} &= \dth{\sigma_{\Sigma, SM}^{RH}} \bigg[ 1 + 
s \frac{C_{ V10}}{M_Z^2 s_Z^2} + f_{\Sigma}^{RH} + 
\mathcal{O} \left( s^{-1} \right) \bigg]                                        
\end{aligned}                                                                 
\end{align}

with

\begin{align}                                                                 
  \begin{aligned}                                                             
    \eta &= \left(1+\cos^2\theta\right)\left(1+8c_Z^4\right)-2\cos\theta      
  \end{aligned}                                                               
\end{align}

It is useful to present the latter results for unpolarized $W$ bosons 
also in a slightly more explicit and complementary form.
In the high-energy limit ($s\gg M^2_W$) the differential cross sections
for the scattering of polarized $e^+e^-$ into unpolarized $W^+W^-$
can be written as
\begin{eqnarray}\label{selww}                                              
\frac{d\sigma(e^-_Le^+_R\to W^-W^+)}{d\cos\theta} &=&                        
\frac{\pi\alpha^2}{2 s}\left[\frac{1-\cos^2\theta}{16 c^4_Z s^4_Z}+         
\frac{(1+\cos\theta)(1+\cos^2\theta)}{2 s^4_Z (1-\cos\theta)} \right. \\    
&-& \left. \frac{s(1-\cos^2\theta)}{8 M^2_W c^2_Z s^4_Z}                   
\left(\delta\kappa_1 - 2\delta\kappa_c +\delta\kappa_Z                     
-2 s^2_Z(\delta\kappa_2 - \delta\kappa_A +\delta\kappa_Z)\right)\right]     
\nonumber                                                                   
\end{eqnarray}
and
\begin{equation}\label{serww}                                              
\frac{d\sigma(e^-_Re^+_L\to W^-W^+)}{d\cos\theta} =                        
\frac{\pi\alpha^2}{2 s} \frac{M^4_Z}{4M^4_W}(1-\cos^2\theta)                 
\left[1+\frac{2s}{M^2_Z}(\delta\kappa_2-\delta\kappa_A +\delta\kappa_Z)\right]
\end{equation}
Here only the leading terms in $M^2_W/s$ have been kept, both
for the standard model results and for the new physics corrections.  
The latter are expressed in terms of the anomalous contributions to the 
couplings, $\delta\kappa_i\equiv\kappa_i-\kappa_{i,SM}$ defined in
(\ref{tgvgam2}) and (\ref{gfvert}). 
In terms of the Lagrangian coefficients one finds for the
parameters that determine the leading corrections
\begin{equation}\label{kapel}                                               
\delta\kappa_1 - 2\delta\kappa_c +\delta\kappa_Z                           
-2 s^2_Z(\delta\kappa_2 - \delta\kappa_A +\delta\kappa_Z)                    
=-C_{V-} - 2 {\rm Re}~C^e_{V9}                                
\end{equation}
\begin{equation}\label{kaper}                                              
\delta\kappa_2 - \delta\kappa_A +\delta\kappa_Z                             
=\frac{C_{V10}}{2 s^2_Z}                                                    
\end{equation}
in agreement with (\ref{slhsum}) and (\ref{srhsum}).

The (full) energy dependence of the leading-order standard model
cross sections is plotted in Figs.~\ref{fig:lhs} and \ref{fig:rhs},
their angular dependence in Figs.~\ref{fig:lha} and \ref{fig:rha}
(solid lines). For illustration, the typical size of potential, $s$-enhanced 
new physics corrections is also indicated (dashed lines).
The following input parameters have been used:
\begin{equation}
M_W=80.4\,{\rm GeV},\qquad M_Z=91.19\,{\rm GeV},
\qquad G_F=1.166\cdot 10^{-5}\,{\rm GeV}^{-2},\qquad \alpha=1/129
\end{equation}
The sine of the weak mixing angle, in the definition used here, 
is then determined through (\ref{szcz}) to be
\begin{equation}\label{sz2num}
s^2_Z = 0.231
\end{equation} 
In order to display the potential impact of new physics we
include for each cross section the leading ${\cal O}(s)$ corrections,
exemplarily setting the relevant coefficients to
$C_{V-}=C_{V9}=C_{V10}=C_{X1}=C_{X2}=C_{X3}=1/(16\pi^2)$. This value corresponds
to the natural size expected from naive dimensional analysis.
In the plots, all cross sections are normalized to the quantity
\begin{equation}\label{rdef}
R=\frac{4\pi\alpha^2}{3s}
\end{equation}

We add several comments on the results presented above. 
\begin{itemize}
\item 
It is instructive to recall the large-$s$ behaviour of the
cross sections in the standard model. The dominant ones scale
as $1/s$. They are:
\begin{equation}\label{xsecdom}
\sigma^{LH}_{LL},\qquad \sigma^{LH}_{TT},\qquad \sigma^{RH}_{LL}
\end{equation}
The remaining cross sections are subleading at high energies
and scale as
\begin{equation}
\sigma^{LH}_{LT}\sim \frac{1}{s^2},\qquad 
\sigma^{RH}_{LT}\sim \frac{1}{s^2},\qquad \sigma^{RH}_{TT}\sim \frac{1}{s^3}
\end{equation}
\item
The leading sensitivity to new physics comes from the ${\cal O}(s)$
enhanced corrections to the dominant cross sections (\ref{xsecdom}).
It depends on the coefficients
\begin{equation}
\sigma^{LH}_{LL}:\, C_{V-,V9},\qquad \sigma^{LH}_{TT}:\, 0,
\qquad \sigma^{RH}_{LL}:\, C_{V10}
\end{equation}
The fact that $\sigma^{LH}_{TT}$ receives no leading corrections
is clearly visible from Figs. \ref{fig:lhs} and \ref{fig:lha}. This feature 
also implies (Fig. \ref{fig:lhs}) that the large-$s$ enhancement in the
cross section for left-handed electrons into unpolarized
$W^+W^-$ is contributed entirely by the longitudinal $W$ bosons,
even though the transverse $W$ bosons have a larger cross section. 
\item 
The CP odd operators ${\cal O}_{XU4}$, ${\cal O}_{XU5}$, ${\cal O}_{XU6}$ and
${\cal O}_{W2}$ do not contribute to the cross sections considered here.
\item
The triple-$W$ operators in (\ref{owi}) arise only at NNLO
($\sim v^2/(16\pi^2\Lambda^2)$) in the effective Lagrangian. Accordingly, 
their coefficients give only subleading contributions to the cross sections. 
The coefficient $C_{W1}$ (CP even operator) enters the correction terms 
$f^{RH}_{LT}$, $f^{LH}_{TT}$, $f^{LH}_{LT}$, $f^{LH}_{\Sigma}$ as well as
the ${\cal O}(s)$ corrections in $\sigma^{RH}_{TT}$ and $\sigma^{LH}_{LT}$.
In the former case $C_{W1}$ is strongly suppressed by a factor
$M^2_Z/\Lambda^2$. In the latter case the suppression is milder,
by a factor $s/\Lambda^2$. However, this is compensated by the overall
suppression of these cross sections at large $s$,
$\sigma^{RH}_{TT}\sim 1/s^3$ and $\sigma^{LH}_{LT}\sim 1/s^2$.  
Therefore the effect of $C_{W1}$ can be expected to be negligible
in practice \cite{Appelquist:1993ka,Arzt:1994gp}.
\item
For high-precision studies standard-model radiative corrections in 
$e^+e^-\to W^+W^-$, which are neglected here, have to be taken into 
account \cite{Denner:2000bj,Denner:2005fg,Bierweiler:2012kw}.
However, these corrections cannot affect the leading relative
corrections from new physics enhanced by $s/M^2_Z$. 
\item
The expression of anomalous couplings in terms of effective theory
coefficients, (\ref{gvkap1}) through (\ref{kapc12}), is fully general and 
can be used to compute further observables in 
$e^+e^-\to W^+W^-$ \cite{Dawson:1993in,Diehl:1993br,Diehl:2002nj}.
\end{itemize}

%%%%%%%%%%%%%%%%%%%%%%%%%%%%%%%%%%%%%%%%%%%%%%%%%%%%%%%%%%%%%%%%%
%   redundant operators
%%%%%%%%%%%%%%%%%%%%%%%%%%%%%%%%%%%%%%%%%%%%%%%%%%%%%%%%%%%%%%%%%
\section{Redundant operators}
\label{sec:redop}

In addition to the dimension-4 operators in (\ref{oxu}),
built from $B_{\mu\nu}$, $W_{\mu\nu}$ and $U$, further operators of similar 
type can be written down. Those may also be used in describing modified
gauge-boson vertices, but they can always be eliminated by 
appropriate field redefinitions (or, equivalently, using equations of motion) 
in favour of the terms in (\ref{oxu})
\cite{Buchalla:2012qq,Nyffeler:1999ap,Grojean:2006nn}.
In this section we discuss how these redundant operators would enter
the anomalous couplings. We also show explicitly how their effect
can be absorbed into the coefficients of the operators 
already present in our basis. This exercise facilitates the transformation
to a different set of independent operators that one might want to
consider. It also provides a useful consistency check of the
expressions in (\ref{gvkap1}) -- (\ref{kapc12}). 

There are 6 redundant operators that have been considered in the
literature, ${\cal O}_{XUi}$, $i=7,\ldots, 12$, in the notation of
\cite{Buchalla:2012qq}. The 3 CP-violating operators $i=10,\, 11,\, 12$
are trivially related to ${\cal O}_{XUi}$, $i=4,\, 5,\, 6$, in (\ref{oxu})
and we will not discuss them further here.
The first of the remaining operators is   
\begin{equation}\label{oxu7}
{\cal O}_{XU7}=-2ig' B_{\mu\nu} \langle D^\mu U^\dagger D^\nu UT_3\rangle
=-i g' g^2 B^{\mu\nu} W^+_\mu W^-_\nu
\end{equation}
It is related to the other operators, up to a total derivative, as
\begin{equation}\label{oxu7rel}
{\cal O}_{XU7}=\frac{g'^2}{2}B_{\mu\nu}B^{\mu\nu} + g'^2 {\cal O}_{\beta_1}
-{\cal O}_{XU1} - g'^2 {\cal O}_{\psi V7} -2 g'^2 {\cal O}_{\psi V10}
\end{equation}
In writing (\ref{oxu7rel}) we have omitted operators similar to 
${\cal O}_{\psi Vi}$ that involve quark fields. The first term
on the r.h.s. only renormalizes the $B$-field kinetic term
and has no effect on the anomalous couplings (see the discussion
in section~\ref{sec:nlophi} below).
Adding a term $C_{X7}{\cal O}_{XU7}$ to the NLO Lagrangian results in
the following shift in the anomalous couplings
\begin{equation}\label{kapxu7}
\Delta\kappa_Z = -\frac{e^2}{c^2_Z} C_{X7},\qquad
\Delta\kappa_A = \frac{e^2}{s^2_Z} C_{X7}
\end{equation}
All other couplings in (\ref{gvkap1}) -- (\ref{kapc12}) remain
unchanged. According to (\ref{oxu7rel}), an inclusion
of $C_{X7}{\cal O}_{XU7}$ in the Lagrangian is equivalent to shifting
the other coefficients by
\begin{equation}\label{delcxu7}
(\Delta\beta_1,\Delta C_{X1},\Delta C_{V7},\Delta C_{V10}) =
C_{X7}\, (g'^2,-1,-g'^2,-2  g'^2)
\end{equation}  
This reflects the redundancy of ${\cal O}_{XU7}$ and
can be checked explicitly with (\ref{gvkap1}) -- (\ref{kapc12}).

Similar considerations apply to the operator
\begin{equation}\label{oxu8}
{\cal O}_{XU8}=-2ig \langle W_{\mu\nu} D^\mu U D^\nu U^\dagger \rangle
\end{equation}
which is related to the other operators as
\begin{equation}\label{oxu8rel}
{\cal O}_{XU8}=g^2 \langle W_{\mu\nu}W^{\mu\nu}\rangle 
-\frac{g^2}{2}v^2\, \langle D_\mu U^\dagger D^\mu U\rangle 
-{\cal O}_{XU1} - 2g^2 {\cal O}_{\psi V8} 
- g^2({\cal O}_{\psi V9} + {\cal O}_{\psi V9}^\dagger)
\end{equation}
up to total derivatives and contributions with quarks.
The first two terms can be absorbed into the
leading-order Lagrangian and have no effect on the
anomalous couplings. 
A term $C_{X8}{\cal O}_{XU8}$ in the Lagrangian would shift the couplings by 
\begin{equation}\label{kapxu8}
\Delta\kappa_Z = \Delta\kappa_A = g^2 C_{X8}\, ,\qquad
\Delta g^Z_1 =\frac{g^2}{c^2_Z} C_{X8}
\end{equation}
with the remaining couplings in (\ref{gvkap1}) -- (\ref{kapc12}) unchanged.
According to (\ref{oxu8rel}), an inclusion
of $C_{X8}{\cal O}_{XU8}$ in the Lagrangian is equivalent to shifting
the other coefficients by
\begin{equation}\label{delcxu8}
(\Delta C_{X1},\Delta C_{V8},\Delta C_{V9},\Delta\delta_G) =
-C_{X8}\, (1,2g^2,g^2,g^2)
\end{equation}  
as can be checked with (\ref{gvkap1}) -- (\ref{kapc12}).

Finally, 
\begin{equation}\label{oxu9}
{\cal O}_{XU9}=-2ig \langle U^\dagger W_{\mu\nu} U T_3\rangle
\, \langle D^\mu U^\dagger D^\nu U T_3 \rangle
\end{equation}
obeys the relation 
\begin{equation}\label{oxu9rel}
{\cal O}_{XU9}=\frac{g^2}{4} \langle W_{\mu\nu}W^{\mu\nu}\rangle 
-\frac{g^2}{8}v^2\, \langle D_\mu U^\dagger D^\mu U\rangle 
-\frac{g^2}{4}{\cal O}_{\beta_1} - \frac{1}{2}{\cal O}_{XU2} 
- \frac{g^2}{4} ({\cal O}_{\psi V9} + {\cal O}_{\psi V9}^\dagger)
\end{equation}
The direct contribution from $C_{X9}{\cal O}_{XU9}$ reads 
\begin{equation}\label{kapxu9}
\Delta\kappa_Z = \Delta\kappa_A = \frac{g^2}{2} C_{X9}
\end{equation}
which, using (\ref{oxu9rel}), is equivalent to shifting the other 
coefficients by
\begin{equation}\label{delcxu9}
(\Delta\beta_1,\Delta C_{X2},\Delta C_{V9},\Delta\delta_G) =
-\frac{C_{X9}}{4}\, (g^2,2,g^2,g^2)
\end{equation}  
This is again consistent with (\ref{gvkap1}) -- (\ref{kapc12}).

We conclude this section with a discussion of an alternative
operator basis, which includes the triple-gauge operators
${\cal O}_{XU7}$, ${\cal O}_{XU8}$ and ${\cal O}_{XU9}$, while eliminating 
three of the original operators in (\ref{ob1}), (\ref{oxu}) and (\ref{opsiv}).
The choice of these three is in principle arbitrary. We emphasize, however,
that it is not possible in general to eliminate all the gauge-fermion 
operators simultaneously since there are more than three (ten without counting 
flavour structure \cite{Buchalla:2012qq}). Because only three gauge-fermion
operators 
(${\cal O}_{\psi V-}$, ${\cal O}_{\psi V9}+ h.c.$, ${\cal O}_{\psi V10}$) happen to
contribute to $e^+e^-\to W^+W^-$, those may indeed be removed altogether from 
the basis
in this case. Additional gauge-fermion terms will be required when other 
processes are considered, such as $W^+W^-$ production from hadronic
initial states (see section \ref{sec:wwlhc}). Restricting our attention
to $e^+e^-\to W^+W^-$ we may write
\begin{align}\label{leffbasis}
&{\cal L}_{eff, NLO} =
\tilde\beta_1 {\cal O}_{\beta_1}+\tilde C_{X1} {\cal O}_{XU1}
+\tilde C_{X2} {\cal O}_{XU2} +\tilde C_{X7} {\cal O}_{XU7}
+\tilde C_{X8} {\cal O}_{XU8} +\tilde C_{X9} {\cal O}_{XU9}+\ldots\nonumber\\
&=\beta_1 {\cal O}_{\beta_1} + C_{X1} {\cal O}_{XU1}
+C_{X2} {\cal O}_{XU2} + C_{V-} {\cal O}_{\psi V-}
+C_{V9} ({\cal O}_{\psi V9} + h.c.) + 
C_{V10} {\cal O}_{\psi V10}+\ldots
\end{align}
where we disregard gauge-fermion operators other than ${\cal O}_{\psi V-}$,
${\cal O}_{\psi V9}$, ${\cal O}_{\psi V10}$. Further operators that are not
affected by the change of basis are understood to be included but are not
written explicitly. In terms of the coefficients, the transformation from 
one to the other basis in (\ref{leffbasis}) is given by
\begin{align}\label{coeffbasis}
\beta_1 &=\tilde\beta_1 +g'^2 \tilde C_{X7} -\frac{g^2}{4}\tilde C_{X9},\quad
C_{X1} =\tilde C_{X1} - \tilde C_{X7} - \tilde C_{X8},\quad
C_{X2} =\tilde C_{X2} - \frac{1}{2}\tilde C_{X9}\nonumber\\
C_{V-} &=-g'^2 \tilde C_{X7} + g^2 \tilde C_{X8},\quad
C_{V9} =-g^2 \tilde C_{X8} -\frac{g^2}{4} \tilde C_{X9},\quad
C_{V10} =-2g'^2 \tilde C_{X7} 
\end{align}

%%%%%%%%%%%%%%%%%%%%%%%%%%%%%%%%%%%%%%%%%%%%%%%%%%%%%%%%%%%%%%%%%               
%   high-energy limit                                                        
%%%%%%%%%%%%%%%%%%%%%%%%%%%%%%%%%%%%%%%%%%%%%%%%%%%%%%%%%%%%%%%%%               
\section{High-energy limit and the Goldstone boson\\ 
equivalence theorem}
\label{sec:helequiv}

The results of section~\ref{sec:crosec}
show that, despite the sizeable number of operators that pa\-ra\-me\-trize 
new physics effects in $e^+e^-\to W^+W^-$, only 3 of them 
appear in the large-energy limit with a relative enhancement factor $s/v^2$, 
thus introducing potential violations of 
unitarity in the $W^+W^-$ cross-section\footnote{Obviously, such divergences 
are actually cut off at the scale of new physics, where new degrees of 
freedom regulate them. Therefore, such divergences never violate unitarity, 
but rather signal the point where the EFT ceases to be valid.}. 
These unitarity violations are associated with the longitudinal modes of the 
$W$ bosons as can be seen by inspection of our results or, more generally, 
by a straightforward application of the equivalence 
theorem~\cite{Cornwall:1974km,Chanowitz:1985hj}. A general discussion
of the equivalence theorem in the context of chiral Lagrangians can be found 
in \cite{GrosseKnetter:1994yp,Dobado:1994vr}. In this section we will 
rederive the large-$s$ limit of the $e^+e^-\to W^+W^-$ cross-section in a 
more transparent way by working in the Landau gauge, 
where the Goldstone modes $\varphi^{\pm}$ appear explicitly. 

\begin{figure}[t]
\begin{center}
\includegraphics[width=4.7cm]{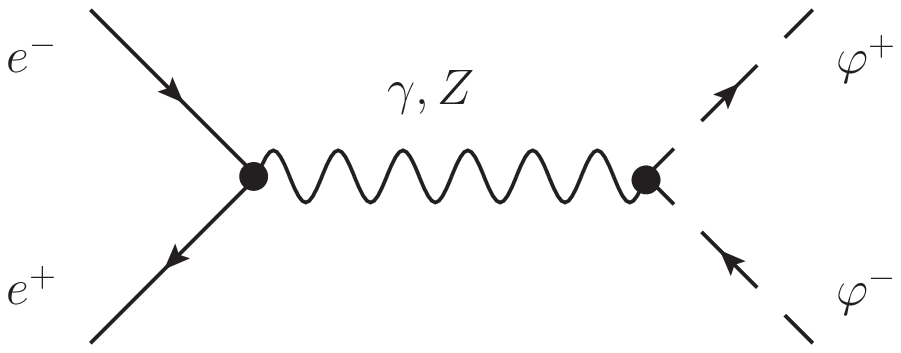}\hskip 0.5cm
\includegraphics[width=3.0cm]{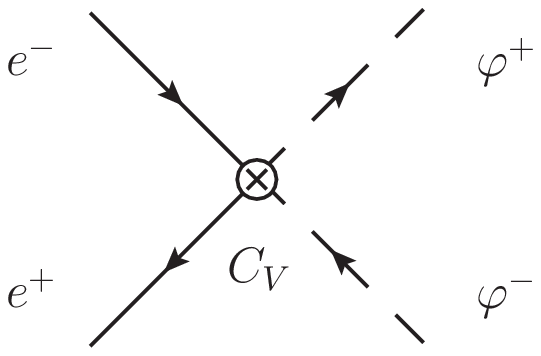}\hskip 0.5cm
\includegraphics[width=4.7cm]{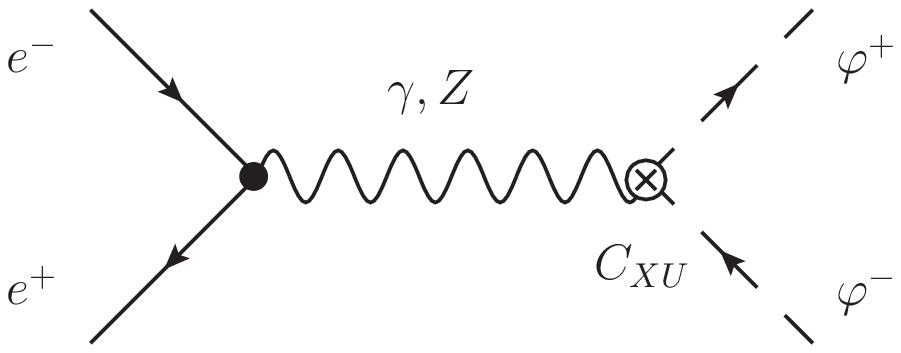}
\end{center}
\caption{\small{\it{Different contributions to $e^+e^-\to \varphi^+\varphi^-$. 
The left-hand diagram is the standard model piece while the central and 
right-hand diagrams are the same contribution from new physics, expressed in 
terms of gauge-fermion (central) or triple-gauge operators (right). $C_{V}$ 
and $C_{XU}$ are short-hand notations for $C_{V7-10}$ and $C_{X7-9}$, 
respectively.}}}\label{fig:4}
\end{figure}

The relevant topologies for $e^+e^-\to \varphi^+\varphi^-$ are collected in 
the second and third dia\-gram of Fig.~\ref{fig:4}. The leftmost diagram is 
the standard model contribution. The $(\gamma,Z)\varphi^+\varphi^-$ vertices 
are obtained from the Goldstone kinetic term
\begin{align}\label{zphism}
\frac{v^2}{4}\langle D^{\mu}U^{\dagger}D_{\mu}U\rangle = 
e\left(\varphi^+i\stackrel{\leftrightarrow}{\partial_{\mu}}\varphi^-\right)
\left(\frac{c^2_Z-s^2_Z}{2 c_Z s_Z}Z^{\mu}+A^{\mu}\right)+\ldots
\end{align}
In the large-$s$ limit, the leading new physics contributions to 
$e^+e^-\to \varphi^+\varphi^-$ can be shown to come only from 
the gauge-fermion operators ${\cal O}_{\psi Vi}$: 
The operator ${\cal O}_{\beta_1}$ contains a $Z\varphi^+\varphi^-$ coupling
proportional to the standard model expression in (\ref{zphism}). This 
contribution is not enhanced in the large-$s$ limit and therefore subleading.
No $(\gamma,Z)\varphi^+\varphi^-$ coupling arises from
${\cal O}_{XU1}$, ${\cal O}_{XU2}$, ${\cal O}_{XU4}$ and ${\cal O}_{XU5}$,
which are bilinear in the gauge fields. Finally, ${\cal O}_{XU3}$ and
${\cal O}_{XU6}$ produce $(\gamma,Z)\varphi^+\varphi^-$ only together with
at least one additional Goldstone particle and therefore do not contribute to
the process of interest here.

The gauge-fermion operators give rise to the 
central diagram in Fig.~\ref{fig:4}. They read explicitly
\begin{align}\label{fermionbasis}
{\cal{O}}_{\psi V7}&=-2{\cal{O}}_{\psi V8}=
\frac{1}{2}({\cal{O}}_{\psi V9} + {\cal{O}}^\dagger_{\psi V9})=
({\bar{e}}_L\gamma^{\mu}e_L)
\frac{1}{v^2}\left(\varphi^+i\stackrel{\leftrightarrow}{\partial_{\mu}}
\varphi^-\right)+\ldots \nonumber\\
{\cal{O}}_{\psi V10}&=({\bar{e}}_R\gamma^{\mu}e_R)\frac{1}{v^2}
\left(\varphi^+i\stackrel{\leftrightarrow}{\partial_{\mu}}\varphi^-\right)
+\ldots
\end{align}
Notice the difference between the unitary and Landau gauge: the gauge-fermion 
operators, which in unitary gauge corrected the $s$ and $t$-channel vertices, 
now take the form of $e^+e^-\varphi^+\varphi^-$ local terms. 

The interference between the standard model and the new physics ($NP$)
contribution can be easily computed and results in
\begin{align}\label{sigcv}
\frac{d\sigma(e^-_R e^+_L\to W^-W^+)_{NP}}{d\cos\theta}
&=\frac{\pi \alpha^2 \sin^2\theta}{8 s_Z^2c_Z^2 M_W^2}C_{V10}\nonumber\\
\frac{d\sigma(e^-_L e^+_R\to W^-W^+)_{NP}}{d\cos\theta}
&=\frac{\pi \alpha^2 \sin^2\theta}{16 s_Z^4 c_Z^2 M_W^2}
\left(C_{V-}+2C_{V9}\right)
\end{align}
which agrees with the results in section~\ref{sec:crosec}
(assuming $C_{V9}$ to be real).

As discussed in section~\ref{sec:redop}, the equations of motion imply 
relations between gauge-fermion, oblique and triple-gauge operators. We have 
already discussed the convenience of working with gauge-fermion operators 
while eliminating triple-gauge operators. However, it is still instructive to 
rederive the large-energy limit in the basis where gauge-fermion operators 
are absent. In this basis, the central diagram in Fig.~\ref{fig:4} gets 
replaced by the rightmost one, where the $(\gamma,Z)\varphi^+\varphi^-$ 
vertices come from the triple-gauge operators
\begin{align}\label{triplebasis}
{\cal{O}}_{XU7}&=-\frac{4ig^{\prime}}{v^2}B_{\mu\nu}
\partial^{\mu}\varphi^+\partial^{\nu}\varphi^-\nonumber\\
{\cal{O}}_{XU8}&=2 {\cal{O}}_{XU9} = -\frac{4ig}{v^2}W^3_{\mu\nu}
\partial^{\mu}\varphi^+\partial^{\nu}\varphi^-
\end{align}  
The results for the cross-sections now take the form
\begin{align}\label{sigcx}
\frac{d\sigma(e^-_R e^+_L\to W^-W^+)_{NP}}{d\cos\theta}
&=-\frac{\pi^2 \alpha^3 \sin^2\theta}{ s_Z^2c_Z^4 M_W^2}C_{X7}
\nonumber\\
\frac{d\sigma(e^-_L e^+_R\to W^-W^+)_{NP}}{d\cos\theta}
&=-\frac{\pi^2 \alpha^3 \sin^2\theta}{4 s_Z^6 c_Z^4 M_W^2}
\left(s_Z^2C_{X7}+c_Z^2\left(C_{X8}+\frac{1}{2}C_{X9}\right)\right)
\end{align}

The equivalence of (\ref{sigcv}) and (\ref{sigcx}) can be checked
using the high-energy version of the equations relating 
${\cal{O}}_{XU7,8,9}$ and ${\cal{O}}_{V7,8,9,10}$ given in 
section~\ref{sec:redop}. In terms of the corresponding coefficients these
relations read
\begin{align}
C_{X7}&=-\frac{c_Z^2}{8\pi\alpha}C_{V10}\nonumber\\
C_{X8}&=\frac{s_Z^2}{4\pi\alpha}\left(C_{V-}-\frac{1}{2}C_{V10}\right)\nonumber\\
C_{X9}&=-\frac{s_Z^2}{\pi\alpha}\left(C_{V-}+C_{V9}-\frac{1}{2}C_{V10}\right)
\end{align}

In the ${\cal O}_{XUi}$ basis (\ref{triplebasis}), the
enhancement $\sim s$ of the  relative corrections is obvious since the
$(\gamma,Z)\varphi^+\varphi^-$ vertices carry three derivatives,
instead of one in the standard model case (\ref{zphism}).
The same enhancement comes about differently in the
${\cal O}_{\psi Vi}$ basis (\ref{fermionbasis}). These operators give
local $e^+e^-\varphi^+\varphi^-$ vertices, which are similar to the 
standard model amplitudes, but without the gauge-boson propagator $\sim 1/s$.
This then leads to the relative enhancement $\sim s$ of the corrections
when they are computed from the ${\cal O}_{\psi Vi}$. 

%%%%%%%%%%%%%%%%%%%%%%%%%%%%%%%%%%%%%%%%%%%%%%%%%%%%%%%%%%%%%%%%%
%   phi-basis
%%%%%%%%%%%%%%%%%%%%%%%%%%%%%%%%%%%%%%%%%%%%%%%%%%%%%%%%%%%%%%%%%
\section{NLO Lagrangian for linearly transforming Higgs}
\label{sec:nlophi}

In the case of a linearly transforming Higgs field, the next-to-leading
order Lagrangian consists of the operators of dimension 5 and 6
listed in \cite{Buchmuller:1985jz,Grzadkowski:2010es}.
The terms of the NLO Lagrangian relevant for $e^+e^-\to W^+W^-$
can be written as 
\begin{equation}\label{lnlozq}
{\cal L}_{\rm NLO} =\frac{1}{\Lambda^2} \sum_{i=1}^{9} z_i Q_i
\end{equation}
with real dimensionless coefficients $z_i$ and the dimension-6
operators
\begin{eqnarray}\label{qq16}
Q_1 &=& (D_\mu \phi^\dagger \phi)(\phi^\dagger D^\mu\phi) \nonumber\\
Q_2 &=& g g' B_{\mu\nu}\, \phi^\dagger W^{\mu\nu} \phi \nonumber\\
Q_3 &=& g g' \varepsilon^{\mu\nu\lambda\rho} 
             B_{\mu\nu}\, \phi^\dagger W_{\lambda\rho} \phi \nonumber\\
Q_4 &=& \bar l\gamma^\mu l\, 
         (\phi^\dagger i D_\mu\phi -iD_\mu\phi^\dagger\phi) \nonumber\\
Q_5 &=& \bar l\gamma^\mu T^a l\, 
         (\phi^\dagger T^a i D_\mu\phi -iD_\mu\phi^\dagger T^a \phi) \nonumber\\
Q_6 &=& \bar e\gamma^\mu e\, 
         (\phi^\dagger i D_\mu\phi -iD_\mu\phi^\dagger\phi) \nonumber\\
Q_7 &=& {\cal O}_{4f}\ ,\quad Q_8 = {\cal O}_{W1}\ ,\quad Q_9 = {\cal O}_{W2}
\end{eqnarray}
We take the Higgs doublet $\phi$ to be normalized such that
its vev is $\langle\phi\rangle =(0,v)^T$ with $v=246\,{\rm GeV}$.

The operators
\begin{equation}\label{qq912}
\begin{array}{ll}
Q_{10} =  \phi^\dagger\phi\, B_{\mu\nu} B^{\mu\nu}\ ,\qquad &
Q_{11} =  \phi^\dagger\phi\, W^a_{\mu\nu} W^{a\mu\nu} \\
Q_{12} =  \phi^\dagger\phi\, \varepsilon^{\mu\nu\lambda\rho} 
                      B_{\mu\nu} B_{\lambda\rho}\ , \qquad &
Q_{13} =  \phi^\dagger\phi\, \varepsilon^{\mu\nu\lambda\rho} 
                      W^a_{\mu\nu} W^a_{\lambda\rho}
\end{array}
\end{equation}
have been omitted from (\ref{lnlozq}) since they have no impact on the 
$e^+e^-\to W^+W^-$ amplitude. This becomes clear in the unitary gauge and 
after dropping contributions with the physical Higgs field $h$, which are of 
no interest in the present case. We may thus replace $\phi^\dagger\phi\to v^2$.  
The operators $Q_{12}$ and $Q_{13}$ then reduce to total derivatives, whereas
$Q_{10}$ and $Q_{11}$ take the form of the usual gauge kinetic terms.
The impact of $Q_{10}$ and $Q_{11}$ 
can be eliminated by a simultaneous rescaling of the gauge field 
and the corresponding gauge coupling \cite{De Rujula:1991se}. 
Ex\-pli\-cit\-ly, the contribution 
from $Q_{11}$ is eliminated, to first order, through the transformations
$W^a_\mu\to (1+\delta_W) W^a_\mu$ and $g\to (1-\delta_W) g$ with
$\delta_W=2 z_{11} v^2/\Lambda^2$ in the leading-order Lagrangian. 
This holds because the field $W^a_\mu$ enters interaction terms in this
Lagrangian only in the combination $gW^a_\mu$.
In particular, the above transformation leaves $gW^a_\mu$ invariant and the 
non-abelian field strength transforms homogeneously 
as $W^a_{\mu\nu}\to (1+\delta_W) W^a_{\mu\nu}$.
A similar transformation removes the impact of $Q_{10}$. 

Comparing with the NLO Lagrangian in the nonlinear realization of the Higgs
sector, in unitary gauge and for $h\to 0$, one finds that the
coefficients in (\ref{lnloco}) are related to $z_1,\, \ldots, z_9$ as 
\begin{equation}\label{relcizi}
\begin{array}{lll}
\beta_1=-z_1\, v^2/\Lambda^2 \qquad & C_{V7}=-2 z_4\, v^2/\Lambda^2 
\qquad & C_{4f}=z_7 \\
C_{X1}=-z_2\, v^2/\Lambda^2  \qquad & C_{V8}=z_5\, v^2/\Lambda^2 
\qquad & C_{W1}=z_8 \\
C_{X4}=-z_3\, v^2/\Lambda^2 \qquad & 
    C_{V9}=\frac{1}{2}z_5\, v^2/\Lambda^2=C^*_{V9}\qquad & C_{W2}=z_9 \\
 & C_{V10}=-2 z_6\, v^2/\Lambda^2 &
\end{array}
\end{equation}
In addition, since the operators ${\cal O}_{XU2}$, ${\cal O}_{XU3}$,
${\cal O}_{XU5}$, ${\cal O}_{XU6}$ correspond to operators of dimension~8
in the linear-Higgs basis \cite{Buchalla:2012qq}, 
at NLO in this basis we may put 
\begin{equation}\label{cx2356}
C_{X2}=C_{X3}=C_{X5}=C_{X6}=0
\end{equation}
The 15 real parameters $\beta_1$. $C_{X1}$, $\ldots$, $C_{X6}$,
$C_{V7}$,  $C_{V8}$, ${\rm Re}~C_{V9}$, ${\rm Im}~C_{V9}$, $C_{V10}$,
$C_{4f}$, $C_{W1}$ and $C_{W2}$ from the nonlinear
Lagrangian thus reduce to the nine real coefficients $z_1$, $\ldots$, $z_9$
in the linear-Higgs basis.

%%%%%%%%%%%%%%%%%%%%%%%%%%%%%%%%%%%%%%%%%%%%%%%%%%%%%%%%%%%%%%%%%              
%   models                                                                  
%%%%%%%%%%%%%%%%%%%%%%%%%%%%%%%%%%%%%%%%%%%%%%%%%%%%%%%%%%%%%%%%%              
\section{Examples of new physics scenarios} 
\label{sec:models}

In previous sections we already commented on the fact that a global 
electroweak fit of the effective theory coefficients does not seem very 
informative, given the strong correlations between them~\cite{Han:2004az}. 
In order to obtain 
an estimate of the size of the coefficients beyond naive dimensional analysis,
it is then useful to resort to different UV completions. In this section we 
will discuss two such scenarios, which affect $e^+e^-\to W^+W^-$ in a 
complementary way, namely UV completions with heavy fermions (constituent 
technicolor) or with heavy vectors ($Z^{\prime}$ models). 
Models with heavy scalars can be shown to affect 
$e^+e^-\to W^+W^-$ only at the loop level and will therefore not be considered. 

\subsection{Constituent technicolor}
\label{subsec:contc}

Constituent technicolor is a very simple model of strongly coupled dynamics 
first introduced in \cite{Appelquist:1993ka}. The model consists of a flavour 
doublet of chiral heavy fermions ${\cal Q}=({\cal U},{\cal D})^T$ with electric 
charges $\pm1/2$ to preserve anomaly cancellation. Since the strong interaction 
between techniquarks is neglected, except for their dynamical mass, it can be 
considered a model for a fourth quark generation. The full Lagrangian can 
then be written as
\begin{align}
{\cal{L}}&={\cal{L}}_{SM}+i{\bar{{\cal{Q}}}}_LD\!\!\!\!\slash~{\cal{Q}}_L+
i{\bar{{\cal{U}}}}_R D\!\!\!\!\slash~{\cal{U}}_R+
i{\bar{{\cal{D}}}}_R D\!\!\!\!\slash~{\cal{D}}_R-
(m_U{\bar{\cal{Q}}}_LUP_+{\cal{U}}_R+m_D{\bar{\cal{Q}}}_LUP_-{\cal{D}}_R+h.c.)
\end{align}
Integrating out the heavy fermions to one loop induces a direct correction 
to the $ZWW$ and $\gamma WW$ vertices but also to gauge boson bilinears. 
One finds~\cite{Appelquist:1993ka}
\renewcommand\arraystretch{2}
\begin{equation}
\begin{array}{ll}
\tilde\beta_1=\displaystyle\frac{4}{v^2}(m_U+m_D)^2\delta^2\xi\qquad & \\
\tilde C_{X1}=-\xi;\qquad & \tilde C_{X7}=-\xi\\
\tilde C_{X2}=-\displaystyle\frac{16}{5}\delta^2\xi;\qquad & \tilde C_{X8}=
-\left(1-\displaystyle\frac{2}{5}\delta^2\right)\xi\\
\tilde C_{X3}=-2\delta \xi;\qquad & 
\tilde C_{X9}=-\displaystyle\frac{28}{5}\delta^2\xi
\end{array}
\end{equation}
where
\begin{align}
\xi&=\frac{N_{TC}}{96\pi^2};\qquad\qquad\qquad \delta=\frac{m_U-m_D}{m_U+m_D}
\end{align} 
Choosing for illustration $N_{TC}=4$, $\delta=1/60$ and $m_U+m_D=3\,{\rm TeV}$, 
one finds that
$\tilde\beta_1\approx 7\cdot 10^{-4}$, 
$\tilde C_{X1}\approx \tilde C_{X7}\approx \tilde C_{X8}\approx -4\cdot 10^{-3}$, 
$\tilde C_{X3}\approx -1\cdot 10^{-4}$, and
$2 \tilde C_{X2}\approx \tilde C_{X9}\approx -7\cdot 10^{-6}$, 
which comply with the naive dimensional estimate $C_i\sim 1/(16\pi^2)$. 
Using (\ref{coeffbasis}) one can trade the triple-gauge  
operators for gauge-fermion vertices. In the basis we have been using 
in this paper we find
\begin{equation}
\begin{array}{ll}
\beta_1=\displaystyle\left[\frac{4}{v^2}(m_U+m_D)^2\delta^2+
e^2\left(\frac{7\delta^2}{5s_Z^2}-\frac{1}{c_Z^2}\right)\right]\xi\qquad & \\
C_{X1}=\displaystyle\left(1-\frac{2}{5}\delta^2\right)\xi; \qquad & 
C_{V-}=\displaystyle e^2\left[\frac{1}{c_Z^2}-\frac{1}{s_Z^2}
\left(1-\frac{2}{5}\delta^2\right)\right]\xi\\
C_{X2}=-\displaystyle\frac{2}{5}\delta^2\xi;\qquad & 
C_{V9}=\displaystyle\frac{e^2}{s_Z^2}(1+\delta^2)\xi\\
C_{X3}=-2\delta \xi;\qquad & C_{V10}=\displaystyle2\frac{e^2}{c_Z^2}\xi
\end{array}
\end{equation}
Doing the same numerical exercise,
$\beta_1\approx 1.6\cdot 10^{-4}$, 
$C_{X1}\approx 4\cdot 10^{-3}$, $C_{X2}\approx -5\cdot 10^{-7}$, 
$C_{X3}\approx -1\cdot 10^{-4}$, and
$C_{V9} \approx -1.4 C_{V-}\approx 1.7 C_{V10}\approx 1.7\cdot 10^{-3}$. 
Two things are worth noticing: (i) the size of the triple gauge operators is 
big enough to invert the sign of $C_{X1}$ in this change of basis,
while $|C_{X1}|$ remains the same; 
(ii) $C_{X4}=C_{X5}=C_{X6}=0$ because constituent technicolor is CP-conserving.  

\subsection{$Z^{\prime}$ models}

We next consider models with a 
$Z^\prime$ \cite{Galison:1983pa,Langacker:2008yv,Langacker:2009su}, 
following the approach developed 
in~\cite{Babu:1997st}. The $Z^{\prime}$ is the gauge boson of a
local $U(1)^\prime$ symmetry and will be assumed to have a mass 
generated through a dynamical mechanism not necessarily related to 
electroweak symmetry breaking. 
Since we are interested in an EFT approach we will not be concerned with the 
dynamical details. Within these assumptions, we will set to zero a bare 
$Z-Z^{\prime}$ mass-mixing term, implying that the Higgs sector of the standard 
model is charged under $U(1)_Y$, but not under $U(1)^\prime$, and 
{\it vice versa} for the Higgs sector of $Z^\prime$. In contrast, a kinetic 
mixing is in general allowed and will be included.

In formulating the $Z^\prime$ model we will use the chiral Lagrangian
description of the standard-model part, as given in (\ref{lsmlo}).
The results can then be interpreted in two different ways.
Either, electroweak symmetry is dynamically broken and the nonlinear
chiral Lagrangian is non-renormalizable with a cutoff $\Lambda$ at
about a few TeV. In this case the $Z^\prime$ mass should be below that scale. 
The limit of interest is $v\ll M_{Z^\prime} < \Lambda$, in which case $Z^\prime$ 
is a light degree of freedom in the chiral Lagrangian, but still heavy enough
in order to be integrated out at the weak scale $v$. Alternatively, we may
consider the conventional renormalizable standard model with the Higgs field
written in polar coordinates, $H\equiv (\tilde\phi, \phi)=(v+h)U$,
and with the physical Higgs scalar $h$ disregarded, since it does not
enter in the applications of interest here. In this case the $Z^\prime$ mass
could be taken to be (much) larger than a few TeV.

The Lagrangian for the $Z^\prime$ model then reads   
\begin{equation}
{\cal{L}}={\cal L}_{SM,U}(\hat B)-
\frac{1}{4}\hat Z_{\mu\nu}^{\prime} \hat Z^{\prime\mu\nu}-
\frac{\sin\chi}{2} \hat Z_{\mu\nu}^{\prime} \hat B^{\mu\nu}
+\frac{\cos^2\chi}{2}  M^2_{Z^\prime} \hat Z_{\mu}^{\prime} \hat Z^{\prime \mu}
-{\hat g}\sum_{j} \hat Y_j {\bar{f}}_j\gamma_{\mu}f_j \hat Z^{\prime\mu}
\end{equation}
${\cal L}_{SM,U}(\hat B)$ is the lowest-order standard model
Lagrangian (\ref{lsmlo}) where the hypercharge gauge field is identified 
with $\hat B$. It is convenient to eliminate the kinetic mixing using
\begin{align}
\left(
\begin{array}{c}
{\hat{B}}_{\mu}\\{\hat{Z}}_{\mu}^{\prime}
\end{array}
\right)=
\left(
\begin{array}{cc}
1 & -\tan\chi\\
0 & 1/\cos\chi
\end{array}
\right)
\left(
\begin{array}{c}
B_{\mu}\\Z_{\mu}^{\prime}
\end{array}
\right)
\end{align}
This field redefinition modifies the $Z^{\prime}$ coupling to fermions and 
generates a coupling between $Z^{\prime}$ and the Goldstone fields. 
The Lagrangian becomes
\begin{align}
{\cal{L}}={\cal L}_{SM,U}(B)
&-\frac{1}{4}Z_{\mu\nu}^{\prime}Z^{\prime\mu\nu}+
\frac{M_{Z^{\prime}}^2}{2}Z_{\mu}^{\prime}Z^{\prime \mu}
+\frac{v^2}{8} g^{\prime 2}\tan^2\chi Z_{\mu}^{\prime}Z^{\prime \mu}\nonumber\\
&-\bigg[\frac{v^2}{2}g^{\prime}\tan\chi\langle U^\dagger iD_\mu U T_3\rangle +
\sum_j{\tilde{g}}_{j}{\bar{f}}_j\gamma_{\mu}f_j\bigg] Z^{\prime\mu}
\end{align}
where
\begin{equation}
{\tilde{g}}_{j}={\hat{g}}\frac{{\hat{Y}}_{j}}{\cos\chi}-
g^{\prime}Y_{j}\tan\chi
\end{equation}
Integrating out the $Z^{\prime}$ at tree level, and expanding to first order
in $1/M^2_{Z^\prime}$, gives the effective Lagrangian
\begin{align}
{\cal{L}}_{eff}&={\cal{L}}_{SM}+\frac{v^4}{8M_{Z^{\prime}}^2}g^{\prime 2}\tan^2\chi
\langle U^\dagger D_\mu U T_3\rangle^2
-\sum_{i,j}\frac{{\tilde g}_i {\tilde g}_j}{2M_{Z^{\prime}}^2}
({\bar{f}}_i\gamma_{\mu}f_i)({\bar{f}}_j\gamma^{\mu}f_j)\nonumber\\
&-\frac{g^{\prime}v^2\tan\chi}{2M_{Z^{\prime}}^2}\sum_{j}{\tilde g}_j
{\bar{f}}_j\gamma_{\mu}f_j \langle U^\dagger i D^\mu U T_3\rangle
\end{align}
For $e^+e^-\to W^+W^-$ the only relevant operators that receive contributions 
are ${\cal{O}}_{\beta_1}$, ${\cal{O}}_{\psi V7}$ and ${\cal{O}}_{\psi V10}$. 
(Here we will not discuss further the renormalization of $G_F$ due to 
the 4-fermion operators, which is a subleading effect at large $s$.)
The coefficients read   
\begin{align}\label{coeffzprime}
\beta_1&=\frac{v^2}{8M_{Z^{\prime}}^2}g^{\prime 2}\tan^2\chi\nonumber\\
C_{V7}&=-\frac{g^{\prime}v^2\tan\chi}{2M_{Z^{\prime}}^2}{\tilde{g}}_{l}\nonumber\\
C_{V10}&=-\frac{g^{\prime}v^2\tan\chi}{2M_{Z^{\prime}}^2}{\tilde{g}}_e
\end{align}
For illustration we choose $M_Z^{\prime}=1$ TeV, $\sin\chi=0.3$, 
$\hat{g}=g^{\prime}$ and $\hat{Y}_l=\hat{Y}_e=-1$. With this choice of 
parameters one finds that 
$\beta_1\approx 0.9\cdot 10^{-4}$, $C_{V7}\approx 1.1\cdot 10^{-3}$
and $C_{V10}\approx 0.9\cdot 10^{-3}$.

The numerical values in this example are similar to those of
sec.~\ref{subsec:contc}. However, whereas the signs of the relevant couplings
in \ref{subsec:contc} are essentially fixed, the signs could be flipped
in the $Z^\prime$ scenario. This would lead to a clear discrimination between
the two models.

For completeness we will also comment on the linear case. Within the same 
assumptions, one can proceed in an analogous way, replacing the kinetic $U$ 
field term by the corresponding term for the linear Higgs model. 
The Lagrangian now takes the form
\begin{align}
{\cal{L}}={\cal L}_{SM,\phi}(B)
&-\frac{1}{4}Z_{\mu\nu}^{\prime}Z^{\prime\mu\nu}+
\frac{M_{Z^{\prime}}^2}{2}Z_{\mu}^{\prime}Z^{\prime \mu}
+\frac{g^{\prime 2}}{8}\tan^2\chi\,\phi^\dagger\phi\, Z_{\mu}^{\prime}Z^{\prime \mu}
\nonumber\\
&+\bigg[\frac{g^{\prime}}{4}\tan\chi 
(\phi^{\dagger}i\stackrel{\leftrightarrow}{D}_{\mu}\phi)
-\sum_j{\tilde{g}}_{j}{\bar{f}}_j\gamma_{\mu}f_j\bigg] Z^{\prime\mu}
\end{align}
Upon integrating out the $Z^\prime$ boson and matching to the linear basis of 
\cite{Grzadkowski:2010es} one obtains the coefficients $z_1$, $z_4$, $z_6$, 
in the notation of section \ref{sec:nlophi}. Their expression in 
terms of (\ref{coeffzprime}) can be inferred from (\ref{relcizi}).

\section{Comments on $W^+W^-$ production at the LHC}
\label{sec:wwlhc}

It is interesting at this point to discuss how the conclusions we have 
reached in our analysis for linear colliders extend to hadron colliders. 
After LEP~\cite{Alcaraz:2006mx}, both 
Tevatron~\cite{Aaltonen:2009aa,Abazov:2009tr,Abazov:2009hk} and 
LHC~\cite{ATLAS:2012mec,CMS:2013} have also studied $W^+W^-$ production and, 
more generally, bounds on triple gauge couplings. The main advantage of a 
hadron collider over a linear one is that one can disentangle the anomalous 
$WWZ$ and $WW\gamma$ contributions by looking at $W\gamma$ 
production~\cite{Dobbs1} and $WZ$ production~\cite{Dobbs2}. $W^+W^-$ is 
afflicted with a larger background and, at least in principle, bounds are 
expected to be less stringent.

A full-fledged analysis of $W^+W^-$ production at the LHC deserves a separate 
paper. 

Here we will content ourselves with commenting on the 
qualitative features one would expect when an effective field theory point of 
view is adopted. For the qualitative approach we are pursuing it will suffice 
to work at the partonic level. The inclusion of parton distribution functions 
(PDFs), which are required in a complete analysis, will not affect our 
conclusions. A recent analysis of $W^+W^-$ production at the LHC, based on a 
subset of the NLO operators in the linear-Higgs scenario, 
can be found in \cite{Degrande:2012wf,Degrande:2013mh}.

At the operator level the only difference between $W^+W^-$ at linear and 
hadron colliders arises in the initial state vertex 
(both in $s$ and $t$ channels), where the hadronic initial 
state has twice the number of operators as the leptonic one. 
To be more precise, while in $e^+e^-$ colliders one finds the 3 combinations
\begin{align}
\frac{1}{2} {\cal{O}}_{\psi V7} - {\cal{O}}_{\psi V8}, \quad
{\cal{O}}_{\psi V9}+{\cal{O}}_{\psi V9}^{\dagger},\quad
{\cal{O}}_{\psi V10},
\end{align} 
in a $pp$ collider 6 operators contribute, namely
\begin{align}
\frac{1}{2} {\cal{O}}_{\psi V1} \pm {\cal{O}}_{\psi V2},\quad
{\cal{O}}_{\psi V3}+{\cal{O}}_{\psi V3}^{\dagger},\quad
{\cal{O}}_{\psi V6}+{\cal{O}}_{\psi V6}^{\dagger},\quad
{\cal{O}}_{\psi V4},\quad
{\cal{O}}_{\psi V5}
\end{align}
The first thing to notice is that while in $e^+e^-\to W^+W^-$ one can trade the 
gauge-fermion operators for triple gauge operators, therefore eliminating 
them altogether, in $pp\to W^+W^-$ this is no longer possible: gauge-fermion 
operators cannot be omitted in general. Obviously 
one can still work in a basis where 3 of the gauge-fermion operators are 
removed. This is however a rather arbitrary choice, which might be sensible 
for a specific process but not for a global electroweak fit. When one is 
interested in fitting more than one process, given the larger number of 
fermions compared to gauge bosons, it seems more natural to remove the triple 
gauge operators instead.   

Even without a detailed analysis one can anticipate the structure of the 
dominant new physics contribution to $pp\to W^+W^-$. Since at $\sqrt{s}=7$~GeV 
the invariant mass of the W pair $\hat{s}$ satisfies 
$M_W^2\ll {\hat{s}}\ll \Lambda^2$, a large-${\hat{s}}$ expansion is warranted. 
Using the equivalence theorem as in section~\ref{sec:helequiv}, 
one can easily conclude 
that 5 out of the 6 gauge-fermion operators contribute at leading-${\hat{s}}$, 
whose precise coefficients can be determined once PDFs are included. 
Therefore, $W^+W^-$ production, somewhat against the common lore, can actually 
be used both at linear and hadron colliders as an excellent probe of new 
physics in the gauge-fermion sector.

%%%%%%%%%%%%%%%%%%%%%%%%%%%%%%%%%%%%%%%%%%%%%%%%%%%%%%%%%%%%%%%%%              
%   conclusions                                                              
%%%%%%%%%%%%%%%%%%%%%%%%%%%%%%%%%%%%%%%%%%%%%%%%%%%%%%%%%%%%%%%%%              
\section{Conclusions}
\label{sec:concl}

In this paper we have analyzed new physics contributions to
the process $e^+e^-\to W^+W^-$, consistently using an
effective field theory treatment. The essential aspects
and results can be summarized as follows:

\begin{itemize}
\item
The analysis employs the most general basis of next-to-leading order 
operators in the electroweak chiral Lagrangian.
\item
Complete relations between the anomalous couplings and
the NLO coefficients in the effective Lagrangian have been derived.
The anomalous couplings include those that modify gauge-fermion interactions.
\item
Equations-of-motion constraints have been discussed and used to eliminate
redundant operators in order to work with a minimal basis of NLO terms.
The redundancy relations imply consistency checks of the relations 
described in the previous item.
\item
Polarized cross sections have been computed for $e^+e^-\to W^+W^-$ with 
both $W$'s on-shell, and with an emphasis on relative corrections 
to first order in the new-physics coefficients.
Specifically, both right- and left-handed electrons, and $W$'s 
with longitudinal ($L$) or transverse ($T$) polarization ($LL$, $LT$, $TT$)
have been considered, as well as the case of an unpolarized $W$ pair.
\item
CP-odd operators do not contribute to the considered observables.
\item
Of particular interest for colliders in the TeV range is the high-energy,
or large-$s$ limit, $M^2_W\ll s\ll\Lambda^2$. The relative corrections 
to the cross sections were quoted explicitly through ${\cal O}(s/M^2_W)$ and 
${\cal O}(1)$ in an $M^2_W/s$ expansion, emphasizing the terms
that grow with $s$.
\item
The relative corrections growing with $s$ have been discussed and explained
with the help of the Goldstone-boson equivalence theorem.
\item
The choice of a basis for the NLO operators is arbitrary in principle
and cannot affect the physics. For illustration we have discussed two 
possible bases and the relation between them. The basis without
redundant triple-gauge boson operators but with all gauge-fermion terms
appears as a convenient choice.
\item
Our results, obtained within the chiral Lagrangian framework,
have also been expressed in terms of the basis of dimension-six operators 
in the standard model with a linearly realized Higgs sector.
The translation is straightforward in the case of $e^+e^-\to W^+W^-$.
\item
The potential size of the new physics coefficients has been estimated using
naive dimensional counting ($C_i\sim 1/16\pi^2$) and 
explicit models (constituent technicolor, $Z'$).
\end{itemize}

The framework discussed here should be useful to identify
and to interpret new physics effects from the dynamics
of electroweak symmetry breaking in studies of $e^+e^-\to W^+W^-$ 
at a TeV-scale linear collider in a systematic way. 
A similar approach can be pursued for many other collider
observables with other final states as well.
Of interest will also be the application to $W$ pair production
at the LHC. Recent measurements \cite{ATLAS:2012mec,CMS:2013}
show somewhat enhanced cross sections for this process. Although the deviation 
from the standard model is not significant at present, such effects
could well be the signature of new physics as described by NLO terms in 
the electroweak effective Lagrangian. The rise with energy of these
effects provides an exciting opportunity, both for the future
running of the LHC at $14\,{\rm TeV}$ and for
$e^+e^-\to W^+W^-$ at a linear collider.

\appendix
%%%%%%%%%%%%%%%%%%%%%%%%%%%%%%%%%%%%%%%%%%%%%%%%%%%%%%%%%%%%%%%%%
%     Appendix
%%%%%%%%%%%%%%%%%%%%%%%%%%%%%%%%%%%%%%%%%%%%%%%%%%%%%%%%%%%%%%%%%

\section{Relative corrections to cross sections\\ 
not enhanced by $s/M^2_Z$}
\label{sec:fij}

In this appendix we list the constant terms in the relative NLO corrections 
to the various standard-model cross sections, denoted by $f^{LH}_i$, 
$f^{RH}_i$ ($i=LL$, $LT$, $TT$, $\Sigma$) in section~\ref{sec:crosec}. 
These terms are of ${\cal O}(1)$ in the expansion for large 
$s/M^2_Z$ and therefore not enhanced in the large-$s$ limit.
We use the definition
\begin{equation} 
C_G \equiv 2(\beta_1 -\delta_G)
\end{equation}

\begin{align}
  \begin{aligned}
f_{TT}^{LH} &=  -\frac{ C_{W1}}{\Lambda^2}\frac{6 e^2 M_Z^2}{s_Z^2 } 
\frac{\left(1-\cos\theta\right)
\Big[1-\cos\theta\left(1+2 c_Z^2\right)\Big]}{1+\cos^2\theta}  + 
\frac{4 e^2 C_{X1}}{c_Z^2 - s_Z^2} - \frac{2 e^2 C_{X2}}{s_Z^2}\\
& \quad +4 \operatorname{Re} C_{ V9}^e + \frac{2c_Z^2 C_G}{c_Z^2 - s_Z^2}\\
f_{TT}^{RH} &= 0
  \end{aligned}
\end{align}

\begin{align}
  \begin{aligned}
    f_{LL}^{LH}	&= \frac{4 c_Z^2 C_{V-}}{\cos \theta -1} 
\Big[ c_Z^2-s_Z^2+\left( 1-6c_Z^2 \right) \cos \theta \Big] + 
2 c^2_Z C_G -4 e^2 C_{X1} \nonumber\\
    & \quad - 2e^2 \frac{1+ 2 c_Z^2}{s_Z^2} C_{X2} + 
4 \operatorname{Re} C_{ V9}^e \bigg[
    8 c_Z^4 \frac{\cos \theta}{1-\cos \theta} - 
\left(s_Z^2 -c_Z^2\right)^2 \bigg] \\
    f_{LL}^{RH}&=-\frac{6 e^2 C_{X1}}{s_Z^2} -\frac{2 e^2 C_{X2}}{s_Z^2}
  \end{aligned}
\end{align}

\begin{align}
  \begin{aligned}
    f_{LT}^{RH} &= -\frac{e^2 C_{X1}}{s_Z^2}-\frac{e^2 C_{X2}}{s_Z^2}+
\frac{2 e^2 C_{X3} \cos \theta}{s_Z^2 \left( 1 + \cos^2 \theta \right)} + 
\frac{C_{W1}}{\Lambda^2}\frac{6 e^2 c_Z^2 M_Z^2 }{s_Z^2} 
  \end{aligned}
\end{align}

\begin{align}
  \begin{aligned}
    f_{LT}^{LH} &= -\frac{4 c_Z^2 C_{V-}}{(\cos \theta-1) \chi^2} 
\bigg[\cos^5 \theta \left(32 c_Z^6+24 c_Z^4-2\right)+\cos^4 \theta 
\left(56 c_Z^6+12 c_Z^4-6 c_Z^2+1\right) \\
    & \quad +2 \cos^3 \theta \left(-8 c_Z^6+8 c_Z^4+4 c_Z^2-1\right)+
2\cos^2 \theta \left(-32 c_Z^6+12 c_Z^4-2 c_Z^2+1\right) \\
    & \quad -8 c_Z^2 \cos \theta \left(2 c_Z^4+3 c_Z^2-1\right) +
8 c_Z^6-20 c_Z^4+2 c_Z^2+1 \bigg] \\
   & +\frac{2 e^2 C_{X1}}{(\cos \theta-1) \left(c_Z^2- s_Z^2\right) \chi^2} 
\bigg[\cos^5 \theta \left(112 c_Z^8+80 c_Z^6-4 c_Z^2+1\right) \\
    & \quad +6c_Z^2 \cos^4 \theta \left(40 c_Z^6+12 c_Z^4-2 c_Z^2+1\right)+
8 c_Z^2 \cos^3 \theta \left(4 c_Z^6+4 c_Z^4-c_Z^2-1\right) \\
    & \quad +4 c_Z^2 \cos^2 \theta \left(-56 c_Z^6+4 c_Z^4-10 c_Z^2+1\right)\\
    & \quad +\cos \theta \left(-144 c_Z^8-80 c_Z^6+24 c_Z^4-4c_Z^2-1\right) -
2 c_Z^2 \left(8 c_Z^6+28 c_Z^4-10 c_Z^2+1\right)\bigg] \nonumber\\
  \end{aligned}
\end{align}

 \begin{align}
  \begin{aligned}
    &+\frac{2 c_Z^2 e^2 C_{X2}}{(\cos \theta-1) s_Z^2 \chi^2} 
\bigg[\cos^5 \theta \left(8c_Z^6-12c_Z^4-18c_Z^2-5\right) \\
    & \quad -2 \cos^4 \theta \left(4 c_Z^6+24 c_Z^4+9 c_Z^2-1\right)+
4\cos^3 \theta \left(-12 c_Z^6+6 c_Z^4+4c_Z^2-1\right) \\
    & \quad +4 \cos^2 \theta \left(-4 c_Z^6+20 c_Z^4+c_Z^2+1\right)+
\cos \theta \left(40 c_Z^6+4 c_Z^4+18 c_Z^2+1\right) \\
    & \quad +2\left(12 c_Z^6-8 c_Z^4+3 c_Z^2+1\right) \bigg] \nonumber\\
  \end{aligned}
\end{align}

 \begin{align}
  \begin{aligned}
    &+\frac{\left(c_Z^2-s_Z^2\right) e^2 C_{X3}}{(\cos \theta-1) s_Z^2 \chi^2} 
\bigg[3 \cos^4 \theta \left(8 c_Z^6+20 c_Z^4+6 c_Z^2-1\right) \\
    & \quad +12  c_Z^2\cos^3 \theta \left(4 c_Z^4+ 4c_Z^2-1\right) -
4 \cos^2 \theta \left(10 c_Z^4 +2c_Z^2 -1\right) \\
    & \quad -4  c_Z^2\cos \theta \left(12 c_Z^4+8 c_Z^2-1\right) -
24 c_Z^6-4 c_Z^4+6 c_Z^2-1 \bigg] \nonumber\\
  \end{aligned}
\end{align}

 \begin{align}
  \begin{aligned}
    &-\frac{4 \operatorname{Re} C_{ V9}^e }{(\cos \theta-1) \chi^2} 
\bigg[\cos^5 \theta \left(32 c_Z^8 +8 c_Z^6 -12 c_Z^4 -2c_Z^2 +1\right) \\
  & \quad +\cos^4 \theta \left(32 c_Z^8 -32 c_Z^6 -12 c_Z^4 +4 c_Z^2-1\right) \\
  & \quad +2 \cos^3 \theta \left(-32 c_Z^8+24 c_Z^6 +8 c_Z^4 -4c_Z^2+1\right) \\
  & \quad +2 \cos^2 \theta \left(-32 c_Z^8+48 c_Z^6-4 c_Z^4+4 c_Z^2-1\right) \\
  & \quad +\cos \theta \left(32 c_Z^8-24 c_Z^6+28 c_Z^4-6 c_Z^2+1\right)
    +32 c_Z^8 -32 c_Z^6 +4 c_Z^4 +4 c_Z^2-1 \bigg] \nonumber\\
  \end{aligned}
\end{align}

 \begin{align}
  \begin{aligned}
    &-\frac{2 c_Z^2  C_G }{\left(c_Z^2-s_Z^2\right) \chi }  
\Big[ \cos^2 \theta \left(-8 c_Z^4-2 c_Z^2+1\right)-12 c_Z^4 \cos \theta -
4 c_Z^4-2 c_Z^2+1 \Big] \nonumber\\
  \end{aligned}
\end{align}

 \begin{align}
  \begin{aligned}
    &  +\frac{C_{W1}}{\Lambda^2}
  \frac{12 e^2 c_Z^2 M_Z^2}{s_Z^2 \left(\cos\theta-1\right) \chi^2} 
\bigg[ \cos^5 \theta \left(24c_Z^8 +20c_Z^6 +6c_Z^4 +3c_Z^2 +1\right)\\
    & \quad +\cos^4 \theta \left(40c_Z^8 +16c_Z^6 +6c_Z^4 +2c_Z^2 -1\right) +
2\cos^3 \theta \left(-8c_Z^8 +12c_Z^6 -2c_Z^2 +1\right)\\
    & \quad +2\cos^2 \theta \left(-24c_Z^8 +8c_Z^6 -6c_Z^4 +2c_Z^2 -1\right)\\
    & \quad -\cos \theta \left(8c_Z^8 +28c_Z^6 -10c_Z^4 +7c_Z^2 -1\right) +
8c_Z^8 -16c_Z^6 -2c_Z^4+ 2c_Z^2 -1\bigg]\\
  \end{aligned}
\end{align}

where

\begin{equation}
\chi = 1+ \left( 2 c_Z^2 \left(1+ \cos \theta \right) + \cos \theta \right)^2 
\end{equation}

\begin{align}
  \begin{aligned}
    f_{\Sigma}^{LH} &=  -\frac{4  c_Z^2 C_{V-}}{ \eta^2}
\frac{(\cos\theta-1)}{(\cos \theta+1)} \bigg[
    \cos^4 \theta \left(16 c_Z^6+40 c_Z^4-10 c_Z^2-1\right)+
2 \cos^3 \theta \left(16 c_Z^6+1\right) \\
    & \quad +4 c_Z^2 \cos^2 \theta \left(8 c_Z^4+24 c_Z^2+1\right)+
2 \cos \theta \left(16 c_Z^6-1\right)+16 c_Z^6-8 c_Z^4+6 c_Z^2+1 \bigg] \\
 & +\frac{4 e^2 C_{X1}}{\left(\cos\theta +1\right)\left(c_Z^2-s_Z^2\right)\eta}
  \bigg[\cos^3 \theta \left(16 c_Z^4-2 c_Z^2-1\right) \\
    & \quad + \cos^2 \theta \left(8 c_Z^4-2c_Z^2+1\right) +
\cos \theta \left(10 c_Z^2-3\right)+8 c_Z^4-6 c_Z^2+3 \bigg] \\
    & -\frac{2 e^2 C_{X2}}{(\cos \theta+1) s_Z^2 \eta} 
\bigg[-\cos^3 \theta \left(c_Z^2-s_Z^2\right)+\cos^2 \theta 
\left(8 c_Z^4+2 c_Z^2-1\right) \\
 & \quad +\cos\theta \left(16 c_Z^4-6 c_Z^2-1\right)+8 c_Z^4+6 c_Z^2+1\bigg] \\
    & +\frac{8 e^2 C_{X3}}{\eta}  \frac{\left(c_Z^2-s_Z^2\right)}{s_Z^2} 
\frac{(\cos \theta-1)}{(\cos \theta+1)} 
\bigg[ \cos \theta \left(c_Z^2+1\right)+c_Z^2 \bigg] \\
    & +\frac{4 \operatorname{Re} C_{ V9}^e}{(\cos \theta+1) \eta^2} 
\bigg[\cos^5 \theta \left(32 c_Z^8-64 c_Z^6+20 c_Z^4+4c_Z^2-1\right) \\
    & \quad +\cos^4 \theta \left(32 c_Z^8+64 c_Z^6-20 c_Z^4-12 c_Z^2+3\right) \\
    & \quad +2 \cos^3 \theta \left(64 c_Z^8-96 c_Z^6-4 c_Z^4+4 c_Z^2-1\right) \\
    & \quad +2 \cos^2 \theta \left(64 c_Z^8+96 c_Z^6+4c_Z^4+4c_Z^2 -1\right) +
 3 \cos \theta \left(32 c_Z^8-4 c_Z^4-4 c_Z^2+1\right)\\ 
    & \quad +96 c_Z^8+12 c_Z^4+4 c_Z^2-1 \bigg] \\
    &+\frac{2 c_Z^2 C_G}{\left(c_Z^2-s_Z^2\right) \eta} 
 \bigg[ \left(\cos^2 \theta +1\right)
    \left(9 c_Z^4-s_Z^4\right) -2 \cos \theta \left(c_Z^2-s_Z^2\right)\bigg]\\
    & + \frac{C_{W1}}{\Lambda^2} \frac{96 e^2 c_Z^4 M_Z^2 }{s_Z^2\eta}
  \frac{\cos\theta-1}{\cos\theta+1}
  \end{aligned}
\end{align}

where

\begin{equation}
    \eta = \left(1+\cos^2\theta\right)\left(1 + 8 c_Z^4 \right)-2\cos\theta
\end{equation}

\begin{align}
  \begin{aligned}
    f_{\Sigma}^{RH} &=
 -\frac{2 e^2 C_{X1} \left(7 + \cos^2 \theta \right)}{s_Z^2 \sin^2 \theta}-
\frac{2 e^2 C_{X2}}{s_Z^2} - \frac{8 e^2 C_{X3} \cos \theta}{s_Z^2 \sin^2\theta}
  \end{aligned}
\end{align}

%%%%%%%%%%%%%%%%%%%%%%%%%%%%%%%%%%%%%%%%%%%%%%%%%%%%%%%%%%%%%%%%%
%     Acknowledgements
%%%%%%%%%%%%%%%%%%%%%%%%%%%%%%%%%%%%%%%%%%%%%%%%%%%%%%%%%%%%%%%%%
\section*{Acknowledgements}

We thank Sven Heinemeyer and Wolfgang Hollik for useful discussions.
This work was performed in the context of the ERC Advanced Grant
project `FLAVOUR' (267104) and was supported in part by the 
DFG cluster of excellence `Origin and Structure of the Universe'.
The work of M.S. was supported in part by the Joachim Herz Stiftung.

%%%%%%%%%%%%%%%%%%%%%%%%%%%%%%%%%%%%%%%%%%%%%%%%%%%%%%%%%%%%%%%%%
%     References
%%%%%%%%%%%%%%%%%%%%%%%%%%%%%%%%%%%%%%%%%%%%%%%%%%%%%%%%%%%%%%%%%

\end{document}